\documentclass{pasa}%
\usepackage{rotating}
\usepackage{lscape}
\usepackage{url}
\usepackage{natbib}

\def\simlt{\mathrel{\rlap{\lower 3pt\hbox{$\sim$}}\raise 2.0pt\hbox{$<$}}}
\def\simgt{\mathrel{\rlap{\lower 3pt\hbox{$\sim$}} \raise
2.0pt\hbox{$>$}}}
\def\lsim{\,\lower2truept\hbox{${<\atop\hbox{\raise4truept\hbox{$\sim$}}}$}\,}
\def\gsim{\,\lower2truept\hbox{${> \atop\hbox{\raise4truept\hbox{$\sim$}}}$}\,}

\DeclareMathAlphabet{\mathpzc}{OT1}{pzc}{m}{it}

\title[Far-IR spectroscopic surveys]{Origins Space Telescope: predictions for far-IR spectroscopic surveys}

\author[M. Bonato et al.]
{Matteo Bonato$^{1,2}$\thanks{matteo.bonato@inaf.it},
Gianfranco De Zotti$^2$,
David Leisawitz$^3$,
Mattia Negrello$^4$,
Marcella Massardi$^1$,
Ivano Baronchelli$^5$,
Zhen-Yi Cai$^6$,
Charles M. Bradford$^5$,
Alexandra Pope$^7$,
Eric J. Murphy$^8$,
Lee Armus$^5$ and
Asantha Cooray$^{9}$
\affil{$^1$INAF$-$Istituto di Radioastronomia, and Italian ALMA Regional Centre, Via Gobetti 101, I-40129, Bologna, Italy}%
\affil{$^2$INAF, Osservatorio Astronomico di Padova, Vicolo Osservatorio 5, I-35122 Padova, Italy}%
\affil{$^3$NASA Goddard Space Flight Center, 8800 Greenbelt Rd, Greenbelt MD, USA}%
\affil{$^4$School of Physics and Astronomy, Cardiff University, The Parade, Cardiff CF24 3AA, UK}%
\affil{$^5$California Institute of Technology, Pasadena, CA}%
\affil{$^6$CAS Key Laboratory for Research in Galaxies and Cosmology, Department of Astronomy, University of Science and Technology \\of China, Hefei, Anhui 230026, China}%
\affil{$^7$Department of Astronomy, University of Massachusetts Amherst, Amherst, MA}%
\affil{$^8$National Radio Astronomy Observatory, 520 Edgemont Road, Charlottesville, VA 22903, USA}%
\affil{$^9$Department of Physics \& Astronomy, University of California, Irvine, CA 92697, USA}}

\jid{PASA}
\doi{10.1017/pas.\the\year.xxx}
\jyear{\the\year}

\usepackage{aas_macros}
\usepackage[bookmarks=false]{hyperref}
\hypersetup{colorlinks,citecolor=blue,linkcolor=blue,urlcolor=blue}

\hypersetup{draft}

\begin{document}

\begin{frontmatter}
\maketitle

\begin{abstract}
We illustrate the extraordinary potential of the (far-IR) Origins Survey
Spectrometer (OSS) on board the Origins Space Telescope (OST) to address a
variety of open issues on the co-evolution of galaxies and AGNs. We present
predictions for blind surveys, each of 1000\,h, with different mapped areas
(a shallow survey covering an area of 10\,deg$^{2}$ and a deep survey of
1\,deg$^{2}$) and two different concepts of the OST/OSS: with a 5.9\,m
telescope (Concept 2, our reference configuration) and with a 9.1\,m
telescope (Concept 1, previous configuration). In 1000\,h, surveys with the
reference concept will detect from $\sim 1.9 \times 10^{6}$ to $\sim 8.7
\times 10^{6}$ lines from $\sim 4.8 \times 10^{5}$--$2.7 \times 10^{6}$
star-forming galaxies and from $\sim 1.4 \times 10^{4}$ to $\sim 3.8 \times
10^{4}$ lines from $\sim 1.3 \times 10^{4}$--$3.5 \times 10^{4}$ AGNs.
The shallow survey will detect substantially more sources than the
deep one; the advantage of the latter in pushing detections to lower
luminosities/higher redshifts turns out to be quite limited. The OST/OSS
will reach, in the same observing time, line fluxes more than one
order of magnitude fainter than the SPICA/SMI and will cover a much broader
redshift range. In particular it will detect tens of thousands of galaxies
at $z \geq 5$, beyond the reach of that instrument. The polycyclic aromatic
hydrocarbons lines are potentially bright enough to allow the detection of
hundreds of thousands of star-forming galaxies up to $z \sim 8.5$, i.e. all
the way through the re-ionization epoch. The proposed surveys will allow us
to explore the galaxy--AGN co-evolution up to $z\sim 5.5-6$ with very good
statistics. OST Concept 1 does not offer significant advantages for the
scientific goals presented here.
\end{abstract}

\begin{keywords}
galaxies: luminosity function -- galaxies: evolution -- galaxies: active -- galaxies: starburst -- infrared: galaxies
\end{keywords}
\end{frontmatter}

\section{Introduction}\label{sect:intro}

In the last fifteen years observations with the NASA/\textit{Spitzer} and
ESA/\textit{Herschel} space telescopes have driven great progress in our
understanding of galaxy formation and evolution. It is now clear that the bulk
of the super-massive black hole (SMBH) accretion and star formation (SF) in
galaxies occurred in the redshift range $1 \lesssim z \lesssim 3$
(\citealt{Merloni2008}; \citealt{Delvecchio2014}; \citealt{Madau14}), and that
the galaxy evolution, unlike dark matter structure formation,
developed in a ``top-down'' fashion, with the most massive galaxies and Active
Galactic Nuclei (AGNs)  forming first (``cosmic downsizing''). The nearly
ubiquitous presence of SMBHs at the centres of galaxies, and the close
relationship between their masses and the properties of the spheroidal stellar
components (\citealt{FerrareseFord2005}; \citealt{KormendyHo2013}), point to a
strong interaction between the build-up of mass in stars and BH growth.
Nonetheless, the details of these
interactions, their impact on cosmic downsizing, and the physical processes that
govern SF in galaxies are
still largely unclear. 

Since the most active cosmic star-formation phases and the associated AGN
growth are dust enshrouded, far-infrared (FIR) and sub-millimeter observations
are necessary for their detection and astrophysical characterization.
These are major goals of next-generation observatories.

Imaging alone is insufficient to reach these goals. For example, disentangling
AGN and SF contributions via photometric SED decomposition proved to be
extremely challenging even in the presence of a rich amount of optical,
near-infrared and sub-mm  (\textit{Herschel}) data (i.e.
\citealt{Delvecchio2014}; \citealt{Berta13}). FIR/sub-mm spectroscopy adds key
information, as demonstrated by ISO, \textit{Herschel} and \textit{Spitzer},
albeit with limited sensitivity.

A giant leap forward will be made possible by the FIR to sub-mm imaging and
spectroscopic observations performed by the Origins Space Telescope
(OST)\footnote{\url{http://origins.ipac.caltech.edu/}}. The OST is an evolving
concept for the Far-Infrared Surveyor mission, the subject of one of four
science and technology definition studies supported by NASA Headquarters in
preparation for the 2020 Astronomy and Astrophysics Decadal Survey. By
delivering a three order-of-magnitude gain in sensitivity over previous FIR
missions and high angular resolution to mitigate spatial confusion in deep
surveys, the OST is being designed to cover large areas of the sky efficiently,
enabling searches for rare objects at low and high redshifts.

The Origins Survey Spectrometer (OSS) provides OST's FIR spectroscopic
capabilities and enables the execution of wide and deep surveys that will yield
large statistical samples of galaxies up to high $z$. Two design concepts for
OST are summarized by \citet{Leisawitz18}, and the mission will be described in
detail in a report to the US National Academies' Decadal Survey in 2019. The
OSS has six wide-band grating spectrometer modules, which combine to cover the
full FIR spectral range (25--590\,$\mu$m) simultaneously, with sensitivity
close to the astronomical background limit, taking advantage of new far-IR
detector array technologies. Because each grating module couples to a slit on
the sky with of order 100 beams\footnote{A descoped version of OSS offers half
this number of beams on the sky}, and OST is agile, allowing for scans
perpendicular to the slit at up to 60 mas/s, OST will be a powerful
spatial-spectral survey machine.

\begin{figure}
\begin{center}
\vskip-1.3cm
\includegraphics[trim=1cm 0.5cm 1.8cm 0.5cm,clip=true,width=0.49\textwidth]{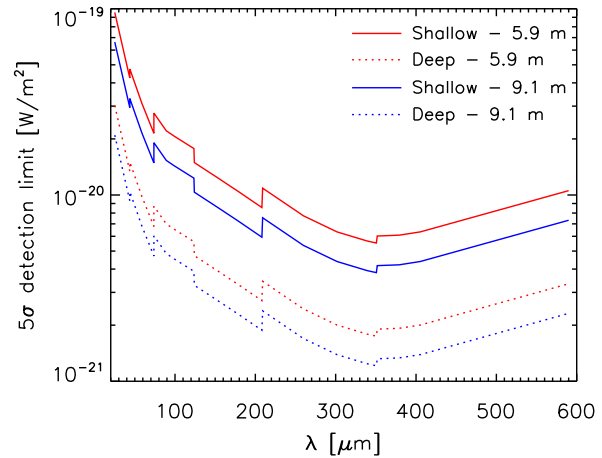}
\vskip-0.5cm
\caption{Detection limits of the OSS instrument on OST as a function of wavelength for
surveys of 1,000\,h each over areas of $1\,\hbox{deg}^2$ (``deep'') and
$10\,\hbox{deg}^2$ (``shallow'') with 5.9\,m and 9.1\,m telescopes.}
 \label{fig:det_lim}
  \end{center}
\end{figure}

Figure~\ref{fig:det_lim} shows the wavelength dependence of the sensitivity  of
OST with the OSS instrument for two studied telescope concepts. The figure
shows $5\,\sigma$ detection limits for surveys of 1,000\,h each over areas of
$1\,\hbox{deg}^2$ and $10\,\hbox{deg}^2$. The wavelength dependence is
primarily attributable to varying background intensity, and secondarily to
instrumental effects. The OST study team considered an off-axis 9.1\,m diameter
telescope (``Concept 1'') and a 5.9\,m diameter on-axis telescope (``Concept
2''). For Concept 2 in survey mode, the mapping speed in units of
[deg$^{2}$(10$^{-19}$\,W\,m$^{-2}$)$^{2}$/sec], with spectral resolving power
$R= 300$, is estimated to be 1.9$\times$10$^{-5}$ (25--44\,$\mu$m),
6.9$\times$10$^{-5}$ (42--74\,$\mu$m), 1.2$\times$10$^{-4}$ (71--124\,$\mu$m),
4.1$\times$10$^{-4}$ (119--208\,$\mu$m), 1.0$\times$10$^{-3}$
(200--350\,$\mu$m) and 7.1$\times$10$^{-4}$ (336--589\,$\mu$m). Mapping speed
is proportional to the telescope's light collecting area, and the area ratio of
the Concept 1 to the Concept 2 telescope is
$52\,\hbox{m}^2/25\,\hbox{m}^2=2.08$. Thus, the mapping speed for Concept 1 is
about twice that of Concept 2. A survey of 1,000\,h over an area of
$1\,\hbox{deg}^2$ with a 5.9\,m telescope implies observing times of 0.58,
1.65, 4.67, 13.1, 37.1 and 106.1\,s per beam at the central wavelengths of the
6 bands listed above, respectively. The observing times per beam at fixed total
time are proportional to the square of the wavelength and inversely
proportional to the total area and to the square of the telescope size.

The OST will perform unprecedented blind FIR/sub-mm spectroscopic surveys deep
and wide enough to provide insight into the physical processes that drive
galaxy evolution out to $z\sim8.5$. In the wavelength interval covered by OSS,
atomic and molecular lines are present over a broad range of excitation levels
for low- and high-$z$ sources. By studying these lines, OST Guest Observers
will be able to derive information about the physical conditions and processes
active  in dust-enshrouded galaxies and AGNs, and measure changes throughout
cosmic history.

The detectable IR spectral lines can come from nuclear activity,
SF regions, or both. Discriminating between the two contributors is not
trivial. However a key diagnostic stems from the fact that AGNs produce harder
radiation and consequently excite metals to higher ionization states than
SF regions. Therefore, lines having high-ionization potentials are a powerful
tool for recognizing nuclear activity \citep{Sturm02,Melendez2008,
GouldingAlexander2009,Weaver2010}, which may be difficult or impossible to identify via
optical spectroscopy in the dusty galaxies with intense SF. Ratios between
pairs of lines with different ionization potentials and comparable critical
densities permit the estimation of the gas ionization state. Ratios between pairs of
lines having comparable ionization potentials and different critical densities
allow us to estimate the gas density (\citealt{Spin92}). \citet{Rubin89}
provides a comprehensive discussion of IR density indicators. \citet{Sturm02}
presented diagnostic diagrams involving different ratios of IR lines (e.g.,
[NeVI]7.63/[NeII]12.81 vs [NeVI]7.63/[OIV]25.89), which can be used to identify composite
(star-forming plus AGN) galaxies and discriminate between the two
components. Other diagnostics exploiting less luminous lines have been
presented by \citet{Voit92} and \citet{Spin92}. Given the large number of
possible combinations of lines, such diagnostic plots are adaptable to
various redshift intervals.

Moreover, the OST with OSS will detect molecular emission lines, in particular CO
lines, which are tracers of  the physical, chemical and dynamical conditions in the
dense gas typically associated with photo-dissociation regions (PDRs) and the
X-ray dominated regions (XDRs) associated to AGNs. The high-J CO lines,
detected by \textit{Herschel} in local galaxies will be observed with the OST
up to high $z$. Such lines, having high ionization potentials, are
indicators of AGN activity. They are also useful tracers of the physics of the
molecular gas. Sub-mm lines are essentially unaffected by dust extinction and
can therefore be used to disentangle the contributions from dust enshrouded
AGNs and star formation in high-redshift galaxies. AGNs and star-forming
galaxies have different low- to high-J CO luminosity ratios (see e.g.
\citealt{CarilliWalter2013}) due to the different excitation mechanisms and
physical conditions (especially temperature and density) in the molecular gas.

In its high-resolution mode, OSS offers velocity resolution
$\lesssim6.2\,\hbox{km}\,\hbox{s}^{-1}$ at 100$\mu$m (\citealt{Bradford18}), and
will provide information on powerful molecular outflows, with velocities of
thousands $\hbox{km}\,\hbox{s}^{-1}$. Such outflows are thought to be driven by
AGN feedback. As is well known, AGN feedback is a key, but still poorly
understood ingredient of most galaxy formation models. A proper study of the
outflow detectability requires dedicated, intensive simulations \citep[cf.,
e.g.,][]{GonzalezAlfonso2017b}, which are beyond the scope of the present
paper.

The OSS on OST will also be able to detect  polycyclic aromatic hydrocarbon
(PAH) spectral features up to very high redshifts. PAH luminosities are useful
indicators of the SF rate \citep[SFR; e.g.][]{Roussel2001,
ForsterSchreiber2004, Peeters2004, Desai2007, Shipley2016}. PAHs were detected
in a large sample of galaxies at quite high $z$ (343 ultra-luminous IR galaxies
at $0.3 < z < 2.8$; \citealt{Kirkpatrick2015}), but it is an open question
whether PAH features exist in galaxies at higher redshifts. There are indeed
indications in the literature of a PAH emission deficit in
low-metallicity/low-luminosity star-forming galaxies compared to
higher-metallicity and/or higher-luminosity SF galaxies (see e.g.
\citealt{Galametz2009}, \citealt{Rosenberg2008}, \citealt{Smith2007},
\citealt{Madden2006}, \citealt{OHalloran2006},
\citealt{Engelbracht2005,Engelbracht2008}).

The mass-metallicity relation (\citealt{Tremonti2004}; \citealt{Zahid2013})
tells us that the metallicity decreases with decreasing stellar mass, and hence
with decreasing luminosity. At high $z$ this effect has a stronger impact since
the average metallicity decreases with increasing $z$ (\citealt{Zahid2013}).
The deficit in PAH emission could be due  to greater permeability of dust
clouds by PAH-destroying or dissociating interstellar radiation
(\citealt{Galliano2003}), as low-metallicity clouds are expected to have lower
dust-to-gas ratios. PAHs are potentially detectable by the OST/OSS up to $z
\sim 8.5$, allowing us - for the first time - to shed light on these issues and
to analyse the properties of the interstellar medium (ISM) in the early
Universe.



Some of the FIR lines detectable by the OST/OSS can be used to measure the
metallicity of galaxies. Since the metal abundance results from the cumulative
star-formation activity and the gas outflow/inflow history in galaxies, the
metallicity of stars and gas in galaxies is an important discriminator between
various galaxy evolutionary scenarios. One of the main IR metallicity
diagnostics is the
([OIII]\,51.81$\mu\hbox{m}$+[OIII]\,88.36$\mu\hbox{m}$)/[NIII]\,57.32$\mu\hbox{m}$
ratio (\citealt{Nagao2011}, \citealt{PereiraSantaella2017}). This diagnostic
tool is applicable to star-forming galaxies and AGNs, and it is almost immune
to extinction and insensitive to temperature, density, and the hardness of the
radiation field. An additional abundance diagnostic is the
([NeII]\,12.81$\mu\hbox{m}$
+[NeIII]\,15.55$\mu\hbox{m}$)/([SIII]\,18.71$\mu\hbox{m}$
+[SIV]\,10.49$\mu\hbox{m}$) ratio (\citealt{FO2016,FO2017}). With OSS, OST
observers will be able to use these diagnostics to measure the metallicities of
hundreds of thousands of galaxies up to z$\sim$6.

In this paper, we adopt the \citet{Cai13} evolutionary model, as
upgraded by \citet{Bonato2014b}. The model was validated by comparison with a
broad array of data: multi-frequency (optical, near-IR, mid-IR, FIR, sub-mm)
luminosity functions of galaxies and AGNs at all available redshifts,
multi-frequency redshift distributions at several flux density limits,
multi-frequency source counts, global and per-redshift slices \citep{Cai13,
Bonato2014a}\footnote{Additional comparisons of model predictions with
observations can be found at \url{http://people.sissa.it/~zcai/galaxy_agn/} or
\url{http://staff.ustc.edu.cn/~zcai/galaxy_agn/index.html}.}. Appendix~\ref{sec:model_data} presents comparisons of
model predictions with more recent data.

Although there are newer models in the literature
\citep[e.g.,][]{Guo2016, Lacey2016, Casey2018}, our model is physically
grounded and has been tested successfully with the broadest variety of observational
data. It also deals self-consistently with the emission of galaxies as a whole,
including both the star-formation and the AGN components, with their dependence
on galaxy age; this is essential for the purposes of the present paper.

This paper is structured as follows. In Section\,\ref{sect:evol} we present the
model. In Section\,\ref{sect:line_vs_IR} we derive correlations between line
and continuum luminosity. In Section\,\ref{sect:results} we describe our
procedure to work out predictions for number counts, IR luminosity functions
and redshift distributions of galaxies and AGNs detectable in the OSS
wavelength range. In Section\,\ref{sect:comparison_concepts} we compare the
capabilities of the OST 5.9\,m and 9.1\,m concepts. In
Section\,\ref{sect:survey_strategy} we investigate in more detail the
expected outcome of surveys of different depth with the 5.9\,m telescope. In
Section\,\ref{sect:discussion} we describe the expected
scientific impact of the OST/OSS surveys, discuss uncertainties, and compare
the OST/OSS performance with that of other FIR/sub-mm instruments. Finally,
in Section\,\ref{sect:summary} we summarize our main conclusions.

Throughout this paper, we adopt a flat $\Lambda \rm CDM$ cosmology
with $\Omega_{\rm m} = 0.31$, $\Omega_{\Lambda} = 0.69$
and $h=H_0/100\, \rm km\,s^{-1}\,Mpc^{-1} = 0.67$
\citep{Planck2015}.

\section{Outline of the model}\label{sect:evol}

In the local Universe, spheroids (i.e. ellipticals and bulges of disk galaxies)
are characterized by relatively old stellar populations with mass-weighted ages
$\gtrsim 8$--9\,Gyr, corresponding to formation redshifts $z\gtrsim 1$--1.5,
while the disk populations are generally younger. Therefore the progenitors of
present day spheroidal galaxies (called proto-spheroids or proto-spheroidal
galaxies) are the dominant star forming population at $z\gtrsim 1.5$, whereas
the SF at $z\lesssim 1.5$ occurs mainly in galaxy disks.

There is clear evidence of a co-evolution of proto-spheroids and of
AGNs hosted by them. Their nuclear activity continues for a relatively short
time, of the order of $10^8\,$yr, after star formation is quenched. At later
times, the central super-massive black holes remain mostly inactive, except for
occasional ``rejuvenations'' due to interactions or mergers. The late nuclear
activity is mostly associated with S0's and spirals, which contain substantial
amounts of interstellar medium that can eventually flow towards the nucleus and
be accreted.

To deal with these different evolutionary paths, our reference model
\citep{Cai13, Bonato2014b} adopts a ``hybrid'' approach. It provides a
physically grounded description of the redshift-dependent co-evolution of the
SFR of spheroidal galaxies and of the accretion rate onto the super-massive
black holes at their centers, while the description of the evolution of
late-type galaxies and of AGNs associated with them is phenomenological and
parametric.

The model considers two sub-populations of late-type galaxies: ``warm''
(starburst) and ``cold'' (normal) galaxies. While in the case of spheroids the
original model by \citet{Cai13} dealt simultaneously with the stellar and
AGN components, AGNs associated with late-type galaxies were evolved as a
separate population. \cite{Bonato2014b} have upgraded the model to allow for co-evolution of
late-type galaxies and their AGNs. This was done by exploiting the mean
relation between SFR and accretion rate derived by \citet{Chen13}, taking into
account the dispersion around it. The relative abundances of type 1 and
type 2 AGNs, as a function of luminosity, were taken into account following
\citet{Hasinger08}. Bright, optically selected QSOs (for which the
\citealt{Chen13} correlation is not applicable) are taken into account by adopting
the best fit evolutionary model by \citet{Croom2009} up to $z=2$ (optical AGNs
at higher $z$ are already included in the \citealt{Cai13} model). This approach
reproduces the observed bolometric luminosity functions of AGNs at different z
(see \citealt{Bonato2014b}).


\begin{table}
\vskip-2.5cm
\caption{Mean values of the log of line-to-IR (8-1000\,$\mu$m) continuum luminosities of galaxies,
$\langle\log({L_{\ell}/L_{\rm IR}})\rangle$, and associated dispersions $\sigma$.
For the PAH\,3.3$\mu$m, PAH\,11.3$\mu$m, PAH\,12.7$\mu$m bands and the H$_{2}$\,17.03$\mu$m, [OI]\,63.18$\mu$m and
[CII]\,157.7$\mu$m  lines, $\langle\log({L_{\ell}/L_{\rm IR}})\rangle$ has been computed
excluding local ULIRGs, for which the line luminosity appears to be uncorrelated
with $L_{\rm IR}$. For the latter objects, the last column gives the mean values of $\log(L_{\ell}/L_\odot)$
and their dispersions.}
\label{tab:c_d_values}
\centering
\footnotesize
\begin{tabular}{lcc}
\hline
\hline
\rule[-3mm]{0mm}{6mm}
Spectral line & $\displaystyle\big\langle\log\big(\displaystyle{L_{\ell}\over L_{\rm IR}}\big)\big\rangle\ (\sigma)$ &
 $\langle \log(L_{\ell})_{\rm UL}/L_\odot\rangle\ (\sigma)$ \\
\hline
${\rm PAH}3.3\mu$m$^{4}$	& -3.11\ (0.30)	& 8.54\	(0.23) \\
${\rm PAH}6.2\mu$m$^{2}$	& -2.20\ (0.36)	&     -   \\
${\rm H_{2}}6.91\mu$m$^{3}$		& -3.97\ (0.39)	&     -   \\
${\rm [ArII]}6.98\mu$m$^{3}$		& -3.96\ (0.32)	&     -   \\
${\rm PAH}7.7\mu$m$^{2}$	& -1.64\ (0.36)	&     -   \\
${\rm PAH}8.6\mu$m$^{2}$	& -2.16\ (0.36)	&     -   \\
${\rm [ArIII]}8.99\mu$m$^{3}$		& -4.22\ (0.69)	&     -   \\
${\rm H_{2}}9.66\mu$m$^{2}$	& -3.96\ (0.52)	&     -   \\
${\rm [SIV]}10.49\mu$m$^{2}$	& -3.95\ (0.69)	&     -   \\
${\rm PAH}11.3\mu$m$^{1}$	& -2.29\ (0.36)	&    9.01\	(0.28)   \\
${\rm H_{2}}12.28\mu$m$^{2}$	& -4.12\ (0.54)	&     -   \\
${\rm HI}12.37\mu$m$^{3}$		& -4.85\ (0.30)	&     -   \\
${\rm PAH}12.7\mu$m$^{4}$	&       -2.20\ (0.33) & 9.75\	(0.28)   \\
${\rm [NeII]}12.81\mu$m$^{1}$	& -3.11\ (0.45)	&     -   \\
${\rm [ClII]}14.38\mu$m$^{3}$		& -5.44\ (0.33)	&     -   \\
${\rm [NeIII]}15.55\mu$m$^{1}$	& -3.69\ (0.47)	&     -   \\
${\rm H_{2}}17.03\mu$m$^{1}$	& -4.04\ (0.46)	&  8.07\	(0.34)   \\
${\rm [FeII]}17.93\mu$m$^{3}$		& -5.18\ (0.34)	&     -   \\
${\rm [SIII]}18.71\mu$m$^{1}$	& -3.49\ (0.48)	&     -   \\
${\rm [ArIII]}21.82\mu$m$^{3}$	& -5.40\ (0.70)	&     -   \\
${\rm [FeIII]}22.90\mu$m$^{3}$	& -6.56\ (0.33)	&     -   \\
${\rm [FeII]}25.98\mu$m$^{3}$		& -4.29\ (0.44)	&     -   \\
${\rm [SIII]}33.48\mu$m$^{1}$	& -3.05\ (0.31)	&     -   \\
${\rm [SiII]}34.82\mu$m$^{1}$	& -2.91\ (0.28)	&     -   \\
${\rm [OIII]}51.81\mu$m$^{1}$        &       -2.84\	(0.44) & -   \\
${\rm [NIII]}57.32\mu$m$^{1}$        &       -3.26\	(0.16) & -   \\
${\rm [OI]}63.18\mu$m        &       -2.93\	(0.30) & 9.01\	(0.30)   \\
${\rm [OIII]}88.36\mu$m      &       -2.92\	(0.48)       &     -   \\
${\rm [NII]}121.90\mu$m$^{1}$        &       -3.49\	(0.36) &     -   \\
${\rm [OI]}145.52\mu$m$^{1}$        &       -3.80\	(0.43) &     -   \\
${\rm [CII]}157.7\mu$m      &        -2.78\	(0.40) & 8.87\	(0.27)   \\
${\rm [NII]} 205.18\mu$m      &        -4.09\	(0.37)       &     -   \\
${\rm [CI]} 370.42\mu$m      &        -4.88\	(0.24)       &     -   \\
${\rm [CI]}	609.14\mu$m      &        -5.14\	(0.33)       &     -   \\
${\rm CO(13-12)}$        &       -5.26\	(0.32)       &     -   \\
${\rm CO(12-11)}$        &       -5.33\	(0.39)       &     -   \\
${\rm CO(11-10)}$        &       -5.23\	(0.33)       &     -   \\
${\rm CO(10-9)}$        &       -5.16\	(0.34)       &     -   \\
${\rm CO(9-8)}$        &       -5.18\	(0.36)       &     -   \\
${\rm CO(8-7)}$        &       -5.12\	(0.35)       &     -   \\
${\rm CO(7-6)}$        &       -5.12\	(0.38)       &     -   \\
${\rm CO(6-5)}$        &       -5.02\	(0.35)       &     -   \\
${\rm CO(5-4)}$        &       -5.02\	(0.34)       &     -   \\
${\rm CO(4-3)}$        &       -5.09\	(0.38)       &     -   \\
\hline
\multicolumn{2}{l}{\scriptsize{$^{1}$Taken from \citet{Bonato2014a}}}\\
\multicolumn{2}{l}{\scriptsize{$^{2}$Taken from \citet{Bonato2014b}}}\\
\multicolumn{2}{l}{\scriptsize{$^{3}$Taken from \citet{Bonato2015}}}\\
\multicolumn{2}{l}{\scriptsize{$^{4}$Taken from \citet{Bonato2017}}}\\
\hline
\hline
\end{tabular}
\end{table}

\begin{figure*}
\begin{center}
\includegraphics[trim=2.6cm 0.25cm 1.4cm 0.4cm,clip=true,width=0.32\textwidth]{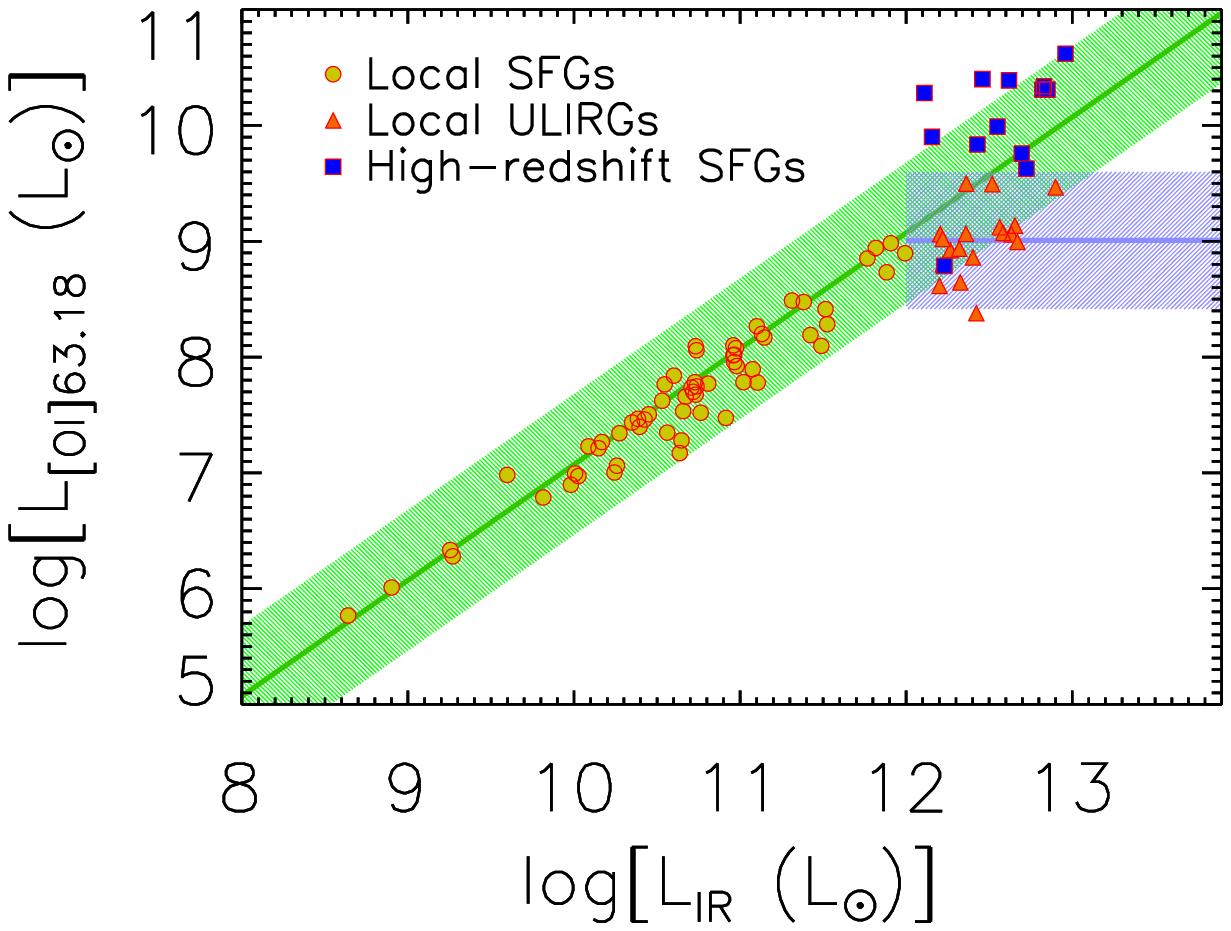}
\includegraphics[trim=2.6cm 0.25cm 1.4cm 0.4cm,clip=true,width=0.32\textwidth]{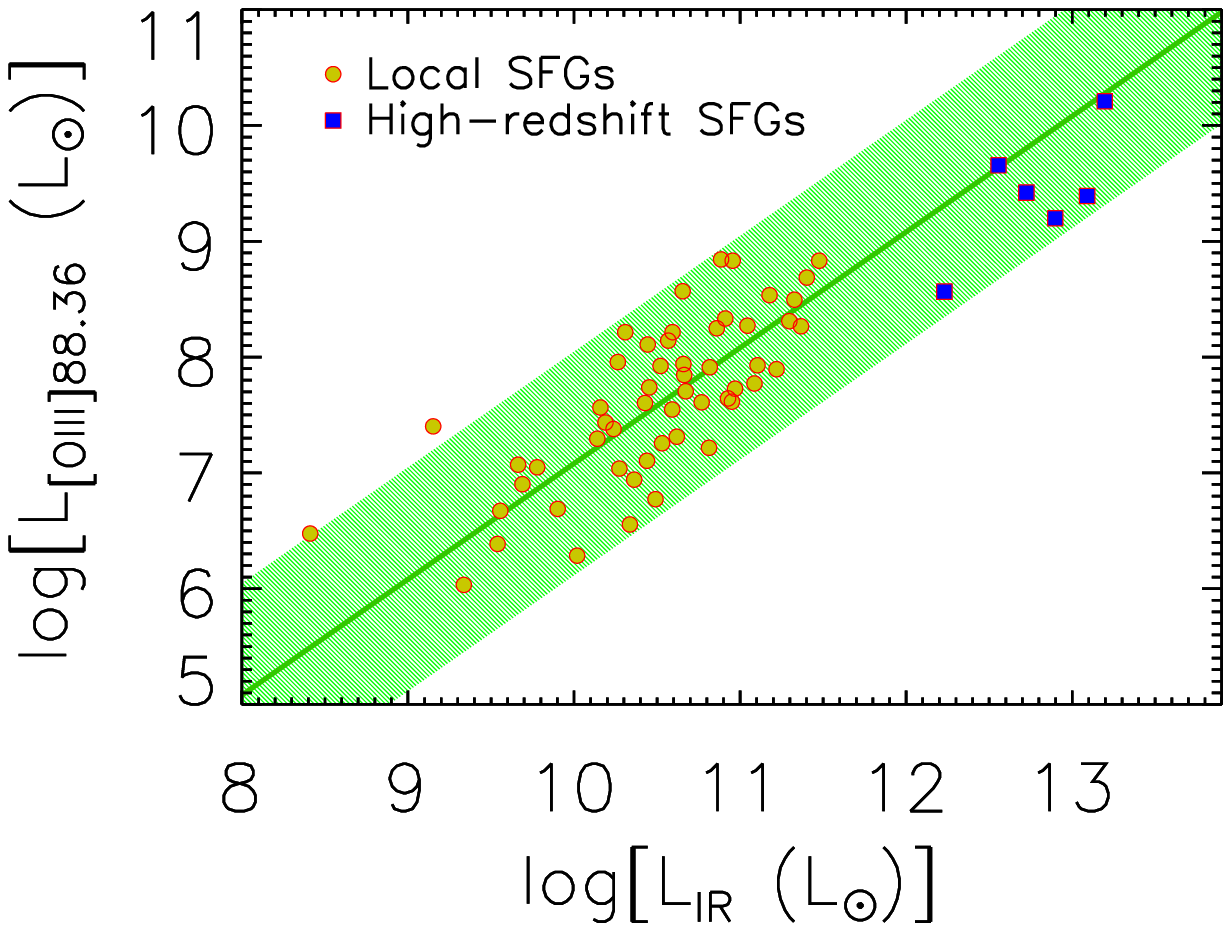}
\includegraphics[trim=2.6cm 0.25cm 1.4cm 0.4cm,clip=true,width=0.32\textwidth]{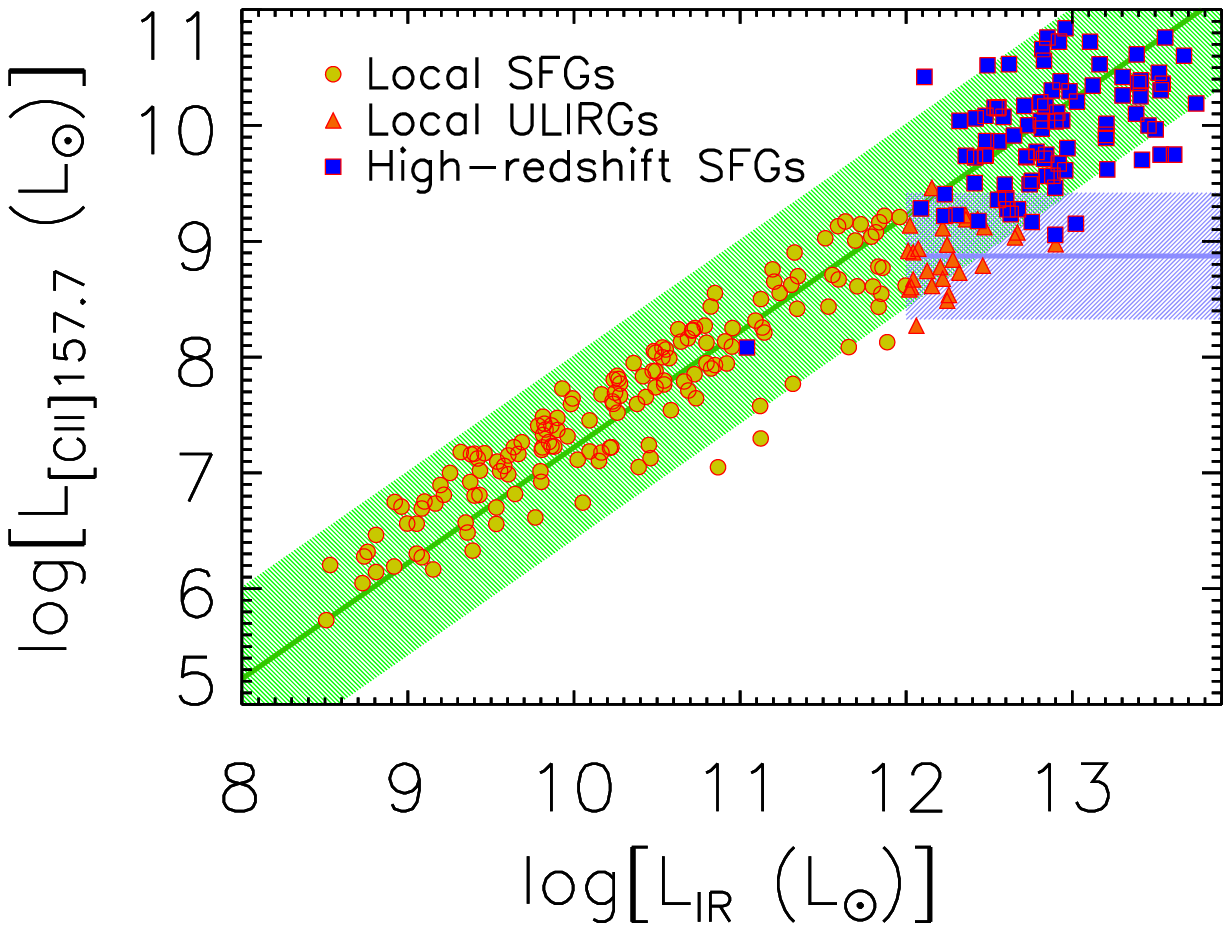}
\caption{Luminosity of the [OI]\,63.18$\mu$m (left panel), [OIII]\,88.36$\mu$m (central panel)
and [CII]\,$157.7\mu$m (right panel) lines, versus continuum IR luminosity. The green bands show the
$2\,\sigma$ range around the mean relation $\log(L_{\ell}/L_\odot)=\log(L_{\rm IR}/L_\odot) + c$ for local SFGs
with $L_{\rm IR}<10^{12}\,L_{\odot}$ (yellow circles) and high-redshift SFGs (blue squares);
the values of $c\equiv\langle \log(L_{\ell}/L_{\rm IR})\rangle$ are given in Table\,\ref{tab:c_d_values}.
The azure bands show the $2\,\sigma$ spread around the mean line luminosity for the sample of local
ULIRGs ($L_{\rm IR}\geq10^{12}\,L_{\odot}$; orange triangles) whose line luminosities appear to be
uncorrelated with $L_{\rm IR}$ and are generally lower than expected from the linear relation holding for
the other sources.  The mean line luminosities $\langle \log (L_{\ell}/L_\odot)\rangle$ of these objects are given in
Table\,\ref{tab:c_d_values} as well. See text for the sources of data points.}
 \label{fig:recal}
  \end{center}
\end{figure*}

\begin{figure*}
\begin{center}
\includegraphics[trim=2.3cm 0.25cm 1.4cm 0.4cm,clip=true,width=0.32\textwidth]{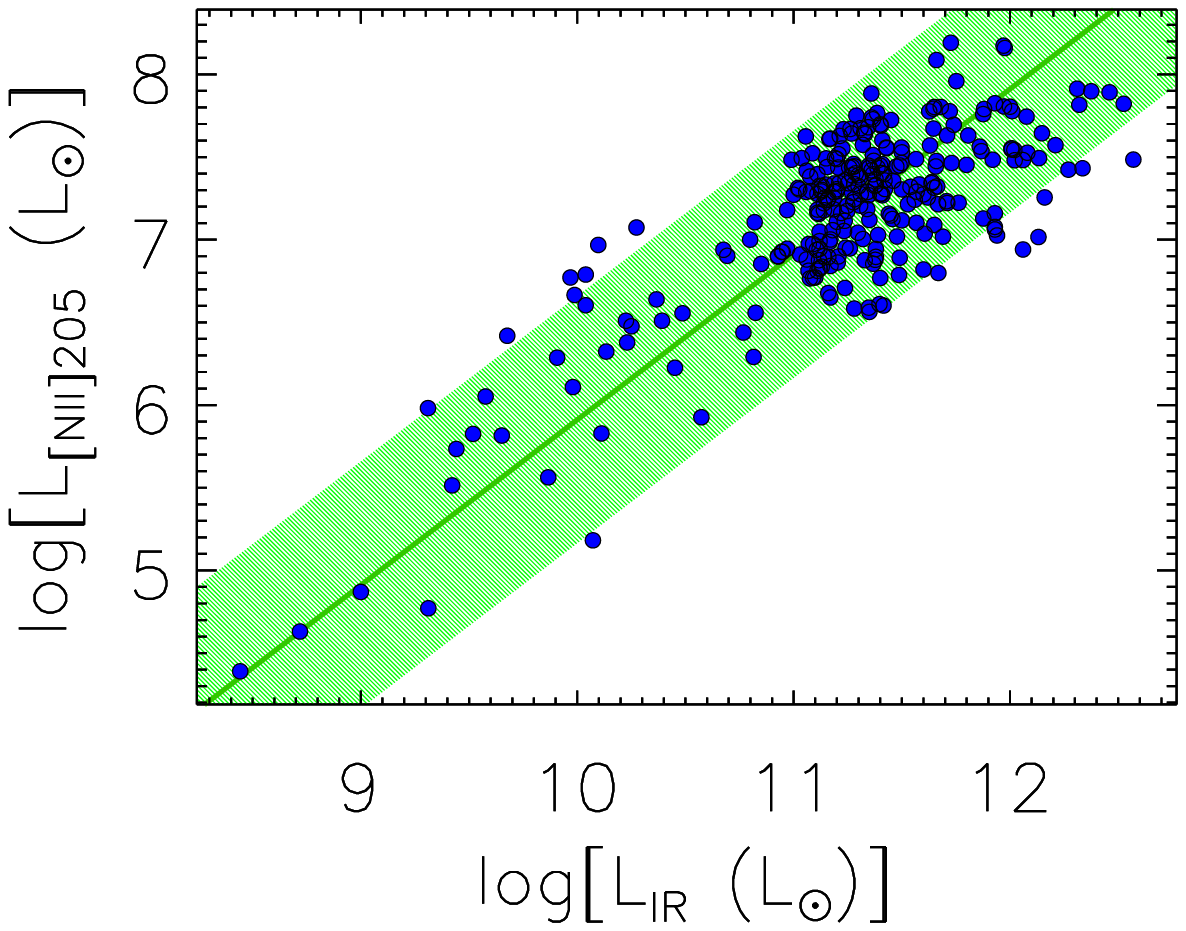}
\includegraphics[trim=2.3cm 0.25cm 1.4cm 0.4cm,clip=true,width=0.32\textwidth]{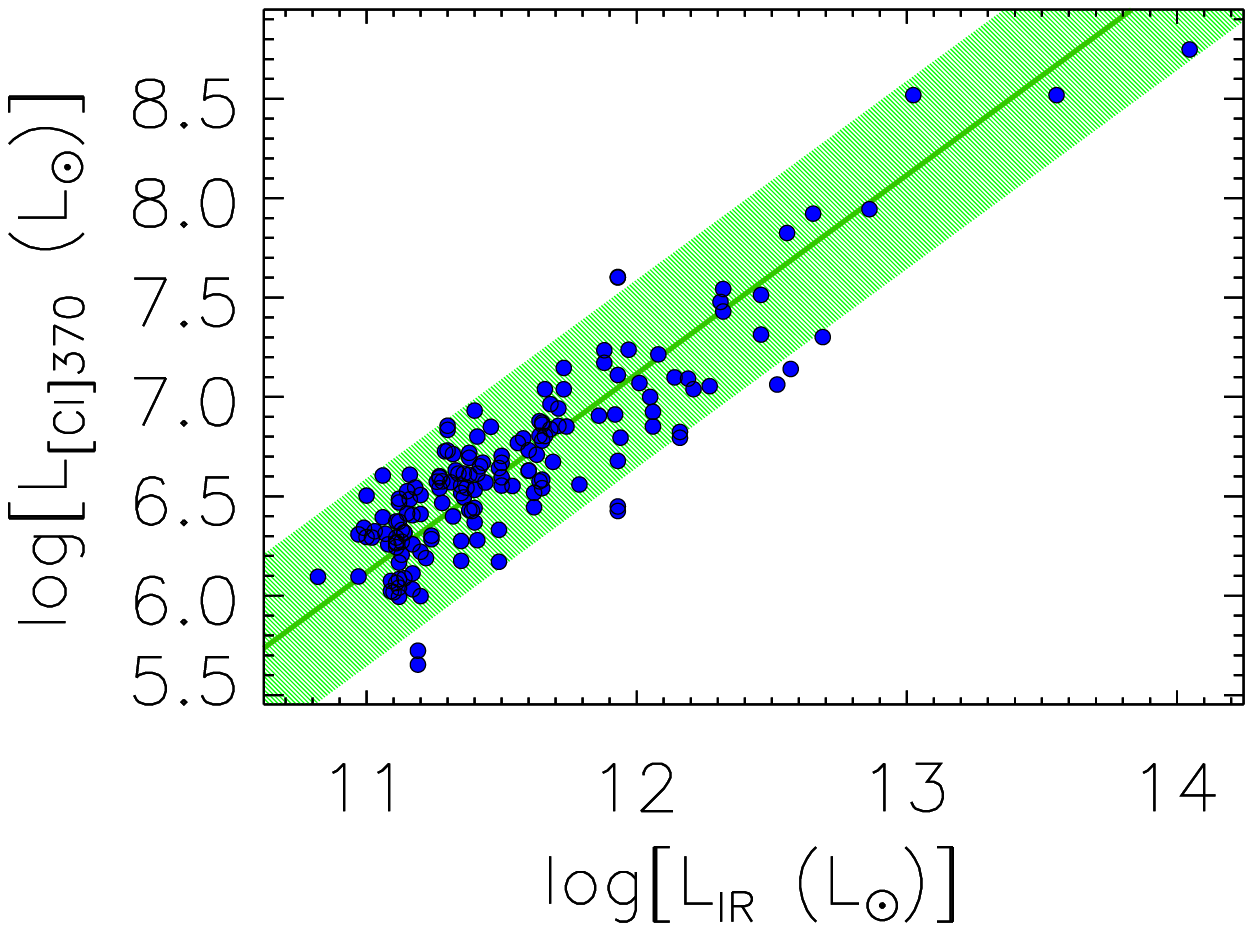}
\includegraphics[trim=2.3cm 0.25cm 1.4cm 0.4cm,clip=true,width=0.32\textwidth]{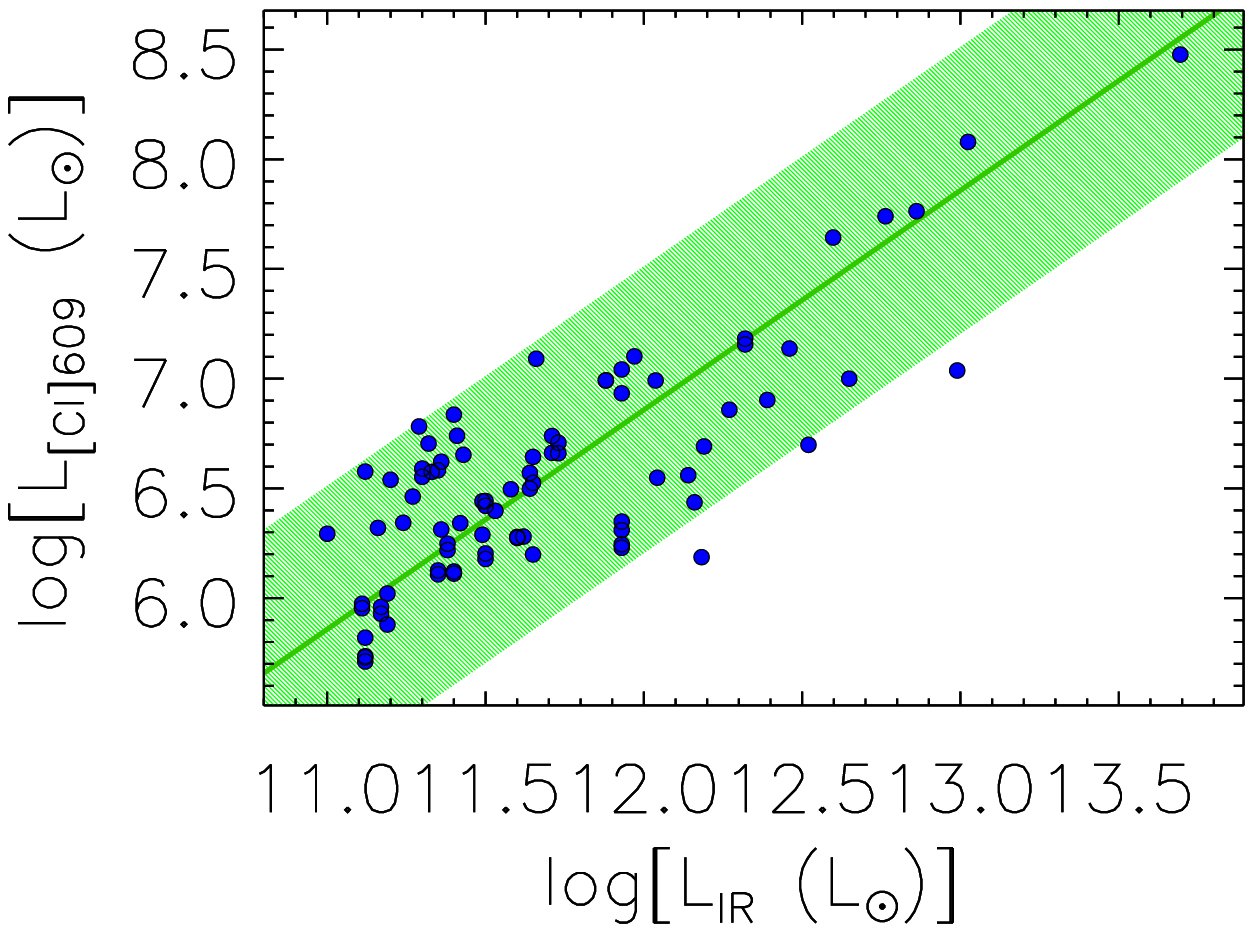}
\caption{Luminosity of the [NII]\,205.2$\mu$m (left panel), [CI]\,370.4$\mu$m (central panel) and
[CI]\,609.1$\mu$m (right panel) lines, versus continuum IR luminosity. The green bands show the $2\,\sigma$ range
around the mean linear relation $\log(L_{\ell}/L_\odot)=\log(L_{\rm IR}/L_\odot) + c$; the values of
$c\equiv\langle \log(L_{\ell}/L_{\rm IR})\rangle$ are given in Table\,\ref{tab:c_d_values}.
See text for the sources of data points.}
 \label{fig:new_cal}
  \end{center}
\end{figure*}


\begin{figure*}
\begin{center}
\includegraphics[trim=4.3cm 4.5cm 4.7cm 1.0cm,clip=true,width=0.99\textwidth]{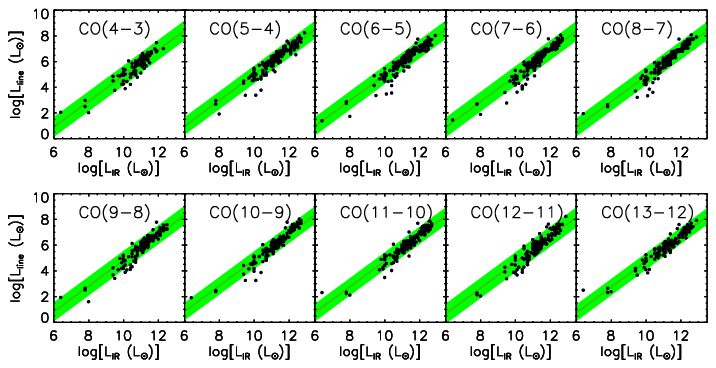}
\caption{Luminosity of the CO J$=$4-3 through J$=$13-12 lines versus continuum IR luminosity. The data
points are from \citet{Kamenetzky2016}. The green band shows the $2\,\sigma$ range around the mean
linear relation $\log(L_{\ell}/L_\odot)=\log(L_{\rm IR}/L_\odot) + c$;
the values of $c\equiv\langle \log(L_{\ell}/L_{\rm IR})\rangle$ are given in Table\,\ref{tab:c_d_values}.}
 \label{fig:CO_cal}
  \end{center}
\end{figure*}

\section{Correlations between continuum and line luminosity}\label{sect:line_vs_IR}

To estimate the number of AGN and galaxy line detections achievable with OST/OSS
surveys, we coupled the SFR/IR luminosity functions (for the galaxies) or the
bolometric luminosity functions (for the AGNs) of each population, as given by
the model, with relationships between line and IR\footnote{SFRs were converted
into $L_{\rm IR}$ following \citet{KennicuttEvans2012}. We neglected the small
differences between the calibration based on the \citet{Kroupa2003} initial
mass function (IMF) used by \citet{KennicuttEvans2012} and that based on the
\citet{Chabrier2003} IMF used by \citet{Cai13}.} or bolometric luminosities.

We used the calibrations derived by
\citet{Bonato2014a,Bonato2014b,Bonato2015,Bonato2017} for 31 IR fine-structure
lines and the 6 polycyclic aromatic hydrocarbon (PAH) bands at 3.3, 6.2, 7.7,
8.6, 11.3 and $12.7\,\mu$m, as they will be accessible in the OST/OSS wavelength range.

The 31 IR fine-structure lines are:
\begin{itemize}
\item 1 coronal region line: [SiVII]\,6.50$\,\mu$m;
\item 8 AGN fine-structure emission lines: [NeVI]\,7.65, [ArV]\,7.90,
    [CaV]\,11.48,  [ArV]\,13.09, [MgV]\,13.50, [NeV]\,14.32, [NeV]\,24.31 and
    [OIV]\,25.89$\,\mu$m;
\item 13 stellar/HII region lines: [ArII]\,6.98, [ArIII]\,8.99, [SIV]\,10.49,
    HI\,12.37, [NeII]\,12.81, [ClII]\,14.38, [NeIII]\,15.55, [SIII]\,18.71,
    [ArIII]\,21.82, [SIII]\,33.48, [OIII]\,51.81, [NIII]\,57.32 and
    [NII]\,121.9$\,\mu$m;
\item 5 lines from photodissociation regions: [FeII]\,17.93, [FeIII]\,22.90,
    [FeII]\,25.98, [SiII]\,34.82 and [OI]\,145.5$\,\mu$m;
\item 4 molecular hydrogen lines: H$_{2}$\,6.91, H$_{2}$\,9.66,
    H$_{2}$\,12.28  and H$_{2}$\,17.03\,$\mu$m.
\end{itemize}

We re-calibrated the [OI]\,63.18 and [CII]\,157.7$\,\mu$m photodissociation
region lines and the [OIII]\,88.36$\,\mu$m stellar/HII region line (whose line
vs. IR luminosity correlations were given in \citealt{Bonato2014a}), and added
more recent data. For these three lines, in addition to the compilation by
\citet{Bonato2014a}, we used data taken from: \citet{George2015}, for all three
lines; \citet{Brisbin2015} and \citet{Farrah2013}, for the [OI]\,63.18 and
[CII]\,157.7$\,\mu$m lines; and \citet{Oteo2016}, \citet{Gullberg2015},
\citet{Schaerer2015}, \citet{Yun2015} and \citet{Magdis2014}, for the
[CII]\,157.7$\,\mu$m line only.  We excluded all objects for which there is
evidence of a substantial AGN contribution, as was done in our previous work.

The line and continuum measurements of strongly lensed galaxies given by
\citet{George2015} were corrected using the gravitational magnifications,
$\mu$, estimated by \citet{Ferkinhoff2014} while those by \citet{Gullberg2015}
were corrected using the magnification estimates from \citet{Hezaveh2013} and
\citet{Spilker2016}, available for 17 out of the 20 sources; for the other
three sources we used the median value $\mu_{\rm median}=7.4$.

For these three lines, Fig.\,\ref{fig:recal} shows the correlations between
line and IR luminosity. In this plot, the galaxies are subdivided into three
groups: local star-forming galaxies (SFGs; $L_{\rm IR}<10^{12}\,L_\odot$,
represented by yellow circles), local Ultra-Luminous IR Galaxies (ULIRGs,
$L_{\rm IR}\geq10^{12}\,L_\odot$, orange triangles), and high-$z$ SFGs (blue
squares).

We warn the reader that the high-$z$ measurements of the three lines
mostly refer to strongly lensed galaxies and are therefore affected by the
substantial and hard to quantify uncertainty on the correction for
magnification, in addition to measurement errors. Not only are estimates of $\mu$
for these objects generally poorly constrained by the available data, but
they do not necessarily apply to the line emitting gas, which may have a
different spatial distribution than the emission used to build the lensing
model \citep[differential lensing;][]{Serjeant2012}. It is, however, reassuring
that high-$z$ galaxies do not substantially deviate, on average, from the
relations defined by the low-$z$ galaxies.

As noted by \citet{Bonato2014a}, luminosities of the [OI]\,63.18 and
[CII]\,157.7$\,\mu$m lines in local ULIRGs do not show any significant
correlation with $L_{\rm IR}$. For such objects we adopted a Gaussian
distribution of the logarithm of the line luminosity, $\log(L_\ell)$, around
its mean value (azure band in Fig.\,\ref{fig:recal}). Line luminosities of the
other populations (low-$z$ non-ULIRGs and high-$z$ SFGs) are consistent with a
direct proportionality between $\log(L_{\ell})$ and $\log(L_{\rm IR})$, i.e.
$\langle\log({L_{\ell}/L_{\rm IR}})\rangle=c\pm\Delta c$, represented by the
green bands in Figs.~\ref{fig:recal}, \ref{fig:new_cal} and
\ref{fig:CO_cal}\footnote{The choice of a direct proportionality between
$\log(L_{\ell})$ and $\log(L_{\rm IR})$ for lines associated with SF is supported
by the extensive simulations (taking into account dust obscuration) carried out
by \citet{Bonato2014a}}. The mean values and the dispersions are listed in
Table\,\ref{tab:c_d_values}.

IR luminosities given over rest-frame wavelength ranges different from the one
adopted here (8--1000\,$\mu$m), were corrected using the following relations,
given by \citet{Stacey2010} and \citet{Gracia-Carpio2008}, respectively:
\begin{eqnarray}
\!\!L_{\rm FIR}(40-500\,\mu{\rm m})\!\!\!\!&=&\!\!\!\!1.5 \times L_{\rm FIR}(42-122\,\mu{\rm m}),\\
\!\!L_{\rm IR}(8-1000\,\mu{\rm m})\!\!\!\!&=&\!\!\!\!1.3  \times L_{\rm FIR}(40-500\,\mu{\rm m}).
\end{eqnarray}
To this initial set of lines we added the [NII]\,205.2$\,\mu$m stellar/HII
region line and the [CI]\,370.4 and [CI]\,609.1$\,\mu$m photo-dissociation
region lines. We collected data on SFGs from: \citet{Lu2017},
\citet{Rosenberg2015} and \citet{Walter2011}  for the [CI]\,370.4 and
[CI]\,609.1$\,\mu$m lines; \citet{Lu2017}, \citet{Herrera-Camus2016} and
\citet{Zhao2013,Zhao2016} for the [NII]\,205.2$\,\mu$m line. In
Fig.\,\ref{fig:new_cal}, we show the correlations between line and IR emission
for these lines. The best-fit values for a direct proportionality and the
associated dispersions are listed in Table\,\ref{tab:c_d_values}.

\begin{table}
  \caption{Coefficients of the best-fit linear relations between line and AGN bolometric luminosities,
  $\log({L_{\ell}}/L_\odot)=a\cdot\log({L_{\rm bol}}/L_\odot)+b$, and 1\,$\sigma$ dispersions around the mean relations.}
  \label{tab:agn_a_b_values}
  \centering
  \footnotesize
  \begin{tabular}{lccc}
    \hline
    \hline
    Spectral line & $a$ & $b$ & disp ($1\sigma$)\\
    \hline
		${\rm [SiVII]}6.50\mu$m$^{2}$		& 0.83	& -3.55	& 0.37	\\
    ${\rm H_{2}}6.91\mu$m$^{2}$		& 0.80	& -2.40	& 0.34	\\
    ${\rm [ArII]}6.98\mu$m$^{2}$		& 0.84	& -4.21	& 0.64	\\
    ${\rm [NeVI]}7.63\mu$m$^{2}$		& 0.79	& -1.48	& 0.42	\\
    ${\rm [ArV]}7.90\mu$m$^{2}$		& 0.87	& -3.85	& 0.32	\\
    ${\rm [ArIII]}8.99\mu$m$^{2}$		& 0.98	& -4.15	& 0.37	\\
    ${\rm H_{2}}9.66\mu$m$^{1}$	& 1.07	& -5.32	& 0.34	\\
    ${\rm [SIV]}10.49\mu$m$^{1}$	& 0.90	& -2.96	& 0.24	\\
    ${\rm [CaV]}11.48\mu$m$^{2}$		& 0.90	& -5.12	& 0.34	\\
    ${\rm H_{2}}12.28\mu$m$^{1}$	& 0.94	& -3.88	& 0.24	\\
    ${\rm HI}12.37\mu$m$^{2}$		& 0.86	& -3.84	& 0.34	\\
    ${\rm [NeII]}12.81\mu$m$^{1}$	& 0.98	& -4.06	& 0.37	\\
    ${\rm [ArV]}13.09\mu$m$^{2}$		& 0.87	& -3.85	& 0.32	\\
    ${\rm [MgV]}13.50\mu$m$^{2}$		& 0.91	& -4.01	& 0.34	\\
    ${\rm [NeV]}14.32\mu$m$^{1}$	& 0.78	& -1.61	& 0.39	\\
    ${\rm [ClII]}14.38\mu$m$^{2}$		& 0.85	& -4.83	& 0.57	\\
    ${\rm [NeIII]}15.55\mu$m$^{1}$	& 0.78	& -1.44	& 0.31	\\
    ${\rm H_{2}}17.03\mu$m$^{1}$	& 1.05	& -5.10	& 0.42	\\
    ${\rm [FeII]}17.93\mu$m$^{2}$		& 0.84	& -3.80	& 0.54	\\
    ${\rm [SIII]}18.71\mu$m$^{1}$	& 0.96	& -3.75	& 0.31	\\
    ${\rm [ArIII]}21.82\mu$m$^{2}$	& 0.98	& -5.34	& 0.36	\\
    ${\rm [FeIII]}22.90\mu$m$^{2}$	& 0.79	& -4.85	& 0.60	\\
    ${\rm [NeV]}24.31\mu$m$^{1}$	& 0.69	& -0.50	& 0.39	\\
    ${\rm [OIV]}25.89\mu$m$^{1}$	& 0.70	& -0.04	& 0.42	\\
    ${\rm [FeII]}25.98\mu$m$^{2}$		& 0.87	& -3.71	& 0.55	\\
    ${\rm [SIII]}33.48\mu$m$^{1}$	& 0.62	& 0.35	& 0.30	\\
    ${\rm [SiII]}34.82\mu$m$^{2}$		& 0.89	& -3.14	& 0.52	\\
		${\rm [OIII]}51.81\mu$m		& 0.92	& -2.47	& 0.32	\\	
    ${\rm [OI]}63.18\mu$m		  & 0.86	& -2.99	& 0.62	\\	
    ${\rm [OIII]}88.36\mu$m		& 0.91	& -3.34	& 0.32	\\	
    ${\rm [OI]}145.52\mu$m		& 0.88	& -4.32	& 0.63	\\	
    ${\rm [CII]}157.70\mu$m		& 0.90	& -4.42	& 0.57	\\
		\hline	
		\multicolumn{2}{l}{\scriptsize{$^{1}$Taken from \citet{Bonato2014b}}}\\
		\multicolumn{2}{l}{\scriptsize{$^{2}$Taken from \citet{Bonato2015}}}\\
    \hline
    \hline
  \end{tabular}
\end{table}

For AGNs, empirical correlations between line and bolometric luminosities
for 11 IR lines in our sample (specifically,  the H$_{2}$\,9.66, [SIV]\,10.49,
H$_{2}$\,12.28, [NeII]\,12.81, [NeV]\,14.32, [NeIII]\,15.55, H$_{2}$\,17.03,
[SIII]\,18.71, [NeV]\,24.31, [OIV]\,25.89 and [SIII]\,33.48$\mu$m lines) were
derived by \citet{Bonato2014b}.

\citet{Bonato2015} used the IDL Tool for Emission-line Ratio Analysis
(ITERA)\footnote{http://home.strw.leidenuniv.nl/~brent/itera.html} to calibrate
the line-to-bolometric luminosity relations for 16 additional AGN IR lines with
insufficient observational data ([SiVII]\,6.50, H$_{2}$\,6.91, [ArII]\,6.98,
[NeVI]\,7.63, [ArV]\,7.90, [ArIII]\,8.99, [CaV]\,11.48, HI\,12.37,
[ArV]\,13.09, [MgV]\,13.50, [ClII]\,14.38, [FeII]\,17.93, [ArIII]\,21.82,
[FeIII]\,22.90, [FeII]\,25.98, [SiII]\,34.82$\mu$m). These calibrations were
found to be in excellent agreement with the available data.

We followed the same procedure to calibrate the line-to-bolometric luminosity
relations for the remaining 10 AGN atomic spectral lines in our sample. The AGN
contributions to the [NIII]\,57.32$\mu$m, [NII]\,121.90/205.18$\mu$m, and
[CI]\,370.42/609.14$\mu$m lines turned out to be negligible. The best-fit
coefficients of the linear relations $\log({L_{\ell}}/L_\odot) =
a\cdot\log({L_{\rm bol}}/L_\odot)+b$ and the associated $1\,\sigma$ dispersions
for the lines in our sample are listed in Table\,\ref{tab:agn_a_b_values}.


\begin{figure*}
  \hspace{+0.0cm} \makebox[\textwidth][c]{
    \includegraphics[trim=2.3cm 7.3cm 6.7cm 2.2cm,clip=true,width=0.99\textwidth,
    angle=0]{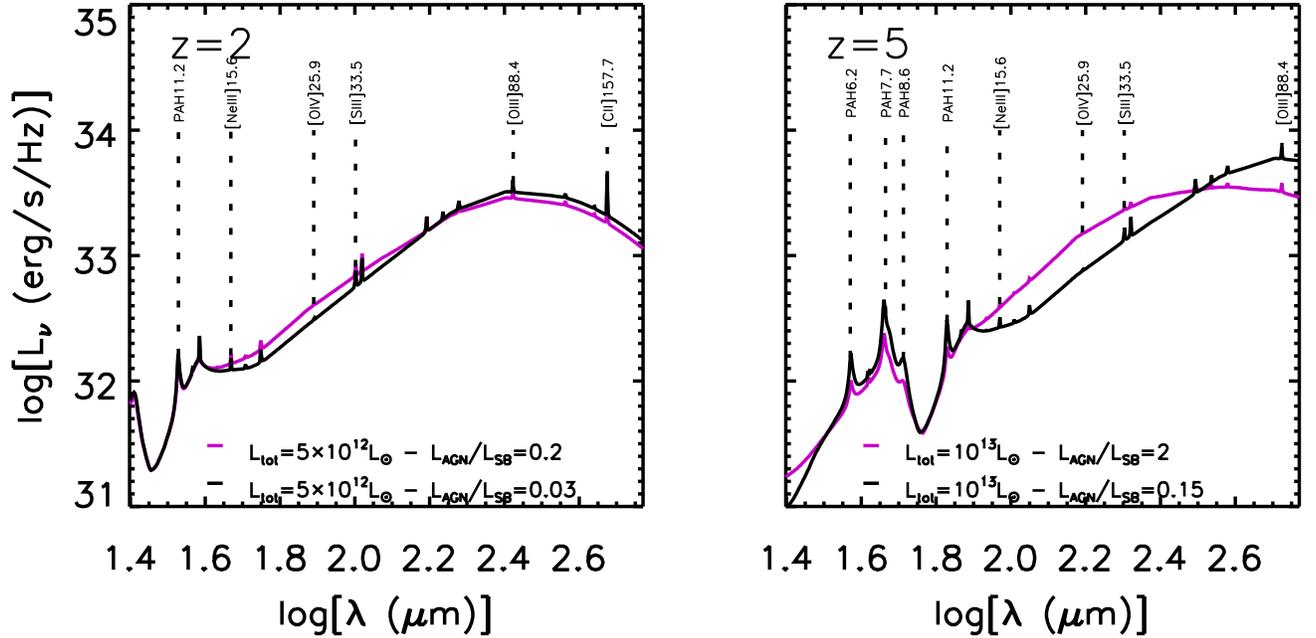}
  }
  \caption{Examples of SEDs with AGN and SF components in the observer frame, over
  the wavelength range of the OST/OSS (25--590$\mu$m). The SEDs are shown for sources at
  $z=2$ (left) and $z=5$ (right) and for the different total source luminosities
  $L_{\rm tot}=L_{\rm AGN}+L_{\rm IR, SF}$  and $L_{\rm AGN}/L_{\rm IR, SF}$ ratios specified in the inset.
  All the spectral lines of our sample are included in the SEDs. Some of the brightest lines are labeled. }
  \label{fig:SED_protsph}
\end{figure*}

Finally, we derived the line-to-IR luminosity relations for 10 CO lines
(J$=$4-3 through J$=$13-12), using the measurements of \citet{Kamenetzky2016}.
For these molecular lines, the data are consistent with both a linear relation
and a direct proportionality between $\log(L_{\rm line})$ and $\log(L_{\rm
IR})$. We  chose a direct proportionality  to avoid an unnecessary second
parameter, as was done for the atomic lines. These relations are shown in
Fig.\,\ref{fig:CO_cal}, with the corresponding best-fit parameter values listed
in Table\,\ref{tab:c_d_values}. {Note that while the AGN contribution to the
high-J CO lines could be significant, the available observational data are not
sufficient to support a derivation of the empirical correlation between AGN
bolometric and CO line luminosity.}

Similarly to what was done in \citet{Bonato2017}, the detection limits for the
broad PAH bands were computed assuming that the spectra are degraded to  $R=50$
from the original $R=300$, resulting in a sensitivity gain by a factor of
$\sqrt{6}$.

\begin{figure}
\centering
    \includegraphics[trim=3.1cm 1.2cm 1.3cm 1.3cm,clip=true,width=0.49\textwidth, angle=0]{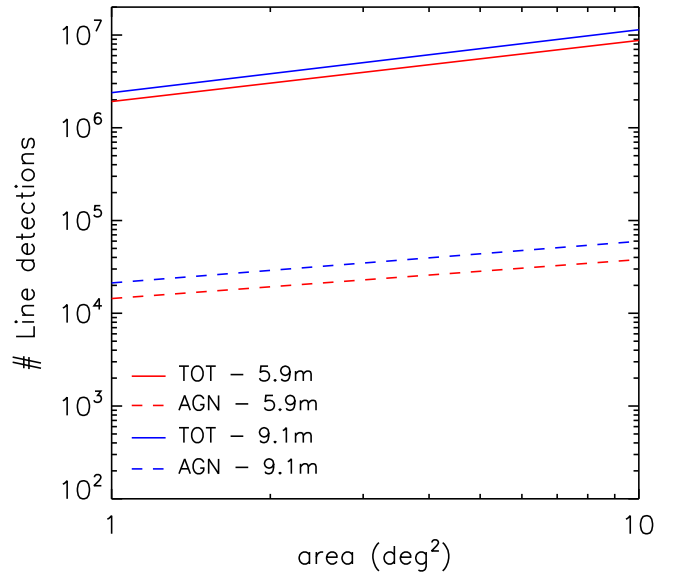}
  \caption{Number of 5$\sigma$ starburst (solid) and AGN (dashed) line detections as a function
  of the mapped area, for a survey of 1000\,h.}
  \label{fig:detections}
\end{figure}

\begin{figure}
\centering
\includegraphics[trim=3.4cm 1.2cm 1.4cm 0.6cm,clip=true,width=0.49\textwidth,
    angle=0]{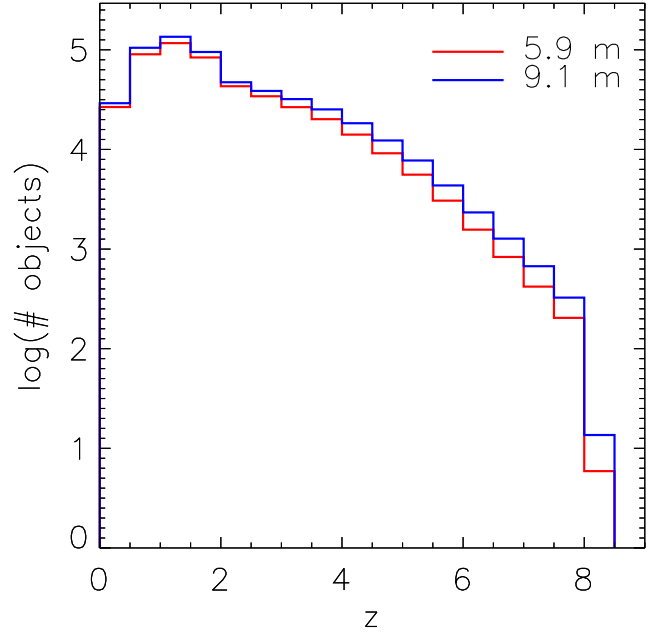}
  \caption{Predicted redshift distributions of galaxies detected
  with a 5.9\,m and a 9.1\,m OST spectroscopic survey in 1000\,h of observing time, assuming an
  areal coverage of 1\,deg$^{2}$.
}
  \label{fig:distributions_sources}
\end{figure}

\begin{figure}
\centering
		\includegraphics[trim=3.4cm 0.9cm 1.4cm 0.6cm,clip=true,width=0.49\textwidth,
     angle=0]{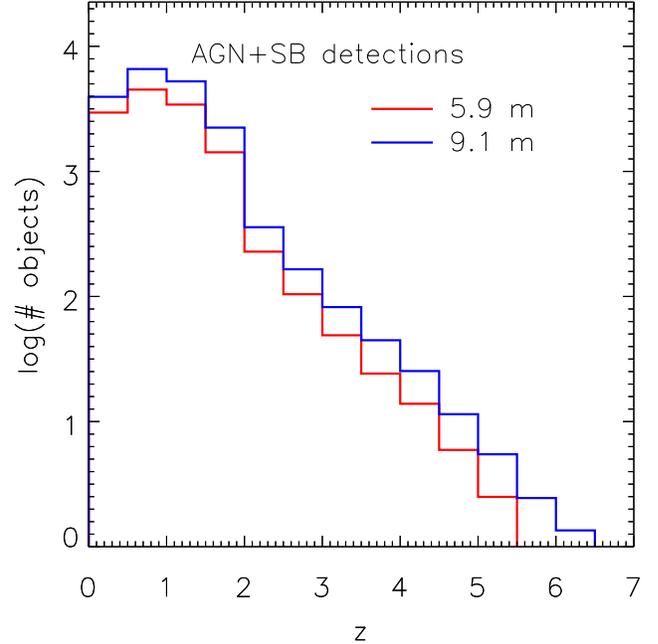}
  \caption{Predicted redshift distributions of galaxies detected
  simultaneously in at least one SF line and one AGN line with a 5.9\,m and a 9.1\,m OST spectroscopic survey in
1000\,h of observing time, assuming an areal coverage of 1\,deg$^{2}$.
  The corresponding AGN redshift distributions are almost identical
  to these, indicating that whenever the AGN component is detected, the SF component is detected as well.}
  \label{fig:distributions_sources_2components}
\end{figure}

\section{Number counts and redshift distributions}\label{sect:results}

We computed redshift-dependent line luminosity functions using the Monte
Carlo approach described in \citet{Bonato2014a, Bonato2014b}. These papers
dealt with dust-obscured star formation phases, for which $L_{\rm
IR}$ is an excellent estimator of the SFR. The line-$L_{\rm IR}$ relations
presented in Section\,\ref{sect:line_vs_IR} were also based on observations of
dusty galaxies, for which the unabsorbed fraction of the UV emission from young
stars is small.

However, the OST with OSS will push the observational frontier to lower
luminosity, higher redshift galaxies, with higher fractions of unabsorbed
UV light, i.e. with smaller fractions of the SFR measured by $L_{\rm
IR}$. The following question then arises: Are the relationships derived above
for the case ``$L_{\rm IR}$ equivalent to star formation luminosity'' to be
interpreted as relations of line luminosity with SFR, or with $L_{\rm
IR}$?

\citet{DeLooze2014} analyzed the reliability of three of the brightest FIR
fine-structure lines, [CII]\,157.7$\,\mu$m [OI]\,63.18 and
[OIII]\,88.36$\,\mu$m, to trace the obscured plus unobscured SFR (measured by
the FIR and by the UV luminosity, respectively) in a sample of low-metallicity
dwarf galaxies from the \textit{Herschel} Dwarf Galaxy Survey. Furthermore, they
extended the analysis to a large sample of galaxies of various types and
metallicities taken from the literature. Their results confirmed that the line
luminosity primarily correlates with the SFR.

We could not find in the literature any other study on the relation between FIR
lines and SFR that took into account both the UV and the FIR
luminosity. In general, the SFR is derived from the FIR luminosity. We have
therefore assumed that the conclusions by \citet{DeLooze2014} can be extended
to all FIR fine structure lines: we do not see any obvious reason why the
luminosity of these lines should depend on the fraction of UV radiation
absorbed by dust.

The situation is different for the PAH emission because these large molecules
may be destroyed by absorption of UV or soft X-ray photons
\citep{Voit1992PAHdestruction}. They more easily survive if they are shielded by
dust. A weakening of PAH emission relative to
dust emission at low gas-phase metallicity was reported by \citet{Engelbracht2005}. \citet{Shipley2016}
found that the PAH luminosity correlates linearly with the SFR, as measured by
the extinction-corrected H$\alpha$ luminosity, for gas-phase metallicity $\log
Z=12+\log(O/H) \ge \log Z_c=8.55$, but strongly decreases with decreasing $Z$
for $Z \le Z_c$.  We have adopted their empirical correction for $Z \le Z_c$
galaxies:
\begin{equation}\label{eq:PAH}
\log L_{\rm PAH, \lambda}^{\rm corr}=\log L_{\rm PAH, \lambda}+ A[\log Z-(\log Z_\odot-\log Z_0)],
\end{equation}
where $L_{\rm PAH, \lambda}$ is the luminosity of the PAH band at the
wavelength $\lambda$ derived from the IR luminosity, taken as a measure of the
SFR (Sect.~\ref{sect:line_vs_IR}), $A$ is given, for different PAH bands, by
Table~4 of \citet{Shipley2016}, and $\log Z_0 = \log Z_\odot-8.55=0.14$. For
the conversion from $L_{\rm IR}$ to SFR we have adopted the calibration by
\citet{KennicuttEvans2012}:
\begin{equation}\label{eq:KE}
\log(\hbox{SFR}/M_\odot\,\hbox{yr}^{-1})=\log(L_{\rm IR}/L_\odot)-9.82.
\end{equation}
A relation between metallicity, SFR, and stellar mass, $M_\star$, approximately
independent of redshift, was found by \citet{Hunt2016}:
\begin{equation}\label{eq:Hunt}
12+\log(O/H)=-0.14 \log(\hbox{SFR})+0.37 \log(M_\star) + 4.82.
\end{equation}
Finally, a redshift-dependent relationship between $M_\star$ and SFR was
derived by \citet{Aversa2015} using the abundance-matching technique. We have
exploited this set of equations to account for the metallicity dependence
of the PAH luminosity.

The molecular lines are even more complex. Two main issues are
at play: the relationship between SFR and molecular gas density
\citep[essentially the Schmidt-Kennicutt law;][]{Kennicutt1998}, whose shape is
still debated, and the uncertain CO-to-$H_2$ conversion factor
\citep{Bolatto2013}. A strong decrease of the CO luminosity to SFR at low
metallicity has long been predicted as a consequence of the enhanced
photo-dissociation of CO by ultraviolet radiation \citep[e.g.,][]{Bolatto2013}.
\citet{Genzel2012} and \citet{Hunt2015} found that a direct proportionality
between $L_{\rm CO}$ and SFR, consistent with that reported in
Section\,\ref{sect:line_vs_IR}, holds for  $Z\gtrsim Z_\odot$, while the
$L_{\rm CO}$/SFR ratio drops for lower metallicity.  We have adopted the
empirical relationship presented by \citet{Hunt2015} for $\log Z\lesssim 8.76$
(see their Fig. 5):
\begin{equation}\label{eq:CO}
\log L_{\rm CO}^{\rm corr}=\log L_{\rm CO}(\hbox{SFR})+ 2.25\log Z-19.71,
\end{equation}
where $L_{\rm CO}(\hbox{SFR})$ is the relationship derived in
Section\,\ref{sect:line_vs_IR}, after converting $L_{\rm IR}$ to SFR using
eq.~(\ref{eq:KE}). The mean metallicity corresponding to a given SFR was
computed as described above.

Clearly, the relationships between the PAH or CO emissions and the SFR (hence
our estimates of the corresponding redshift-dependent luminosity functions and
number counts) are affected by large uncertainties, especially at high
redshifts. FIR spectroscopic surveys conducted with the OST will shed light on these issues.

\begin{figure*}
  \hspace{+0.0cm} \makebox[\textwidth][c]{
    \includegraphics[trim=1.3cm 0.7cm 1.7cm 1.9cm,clip=true,width=0.49\textwidth,
    angle=0]{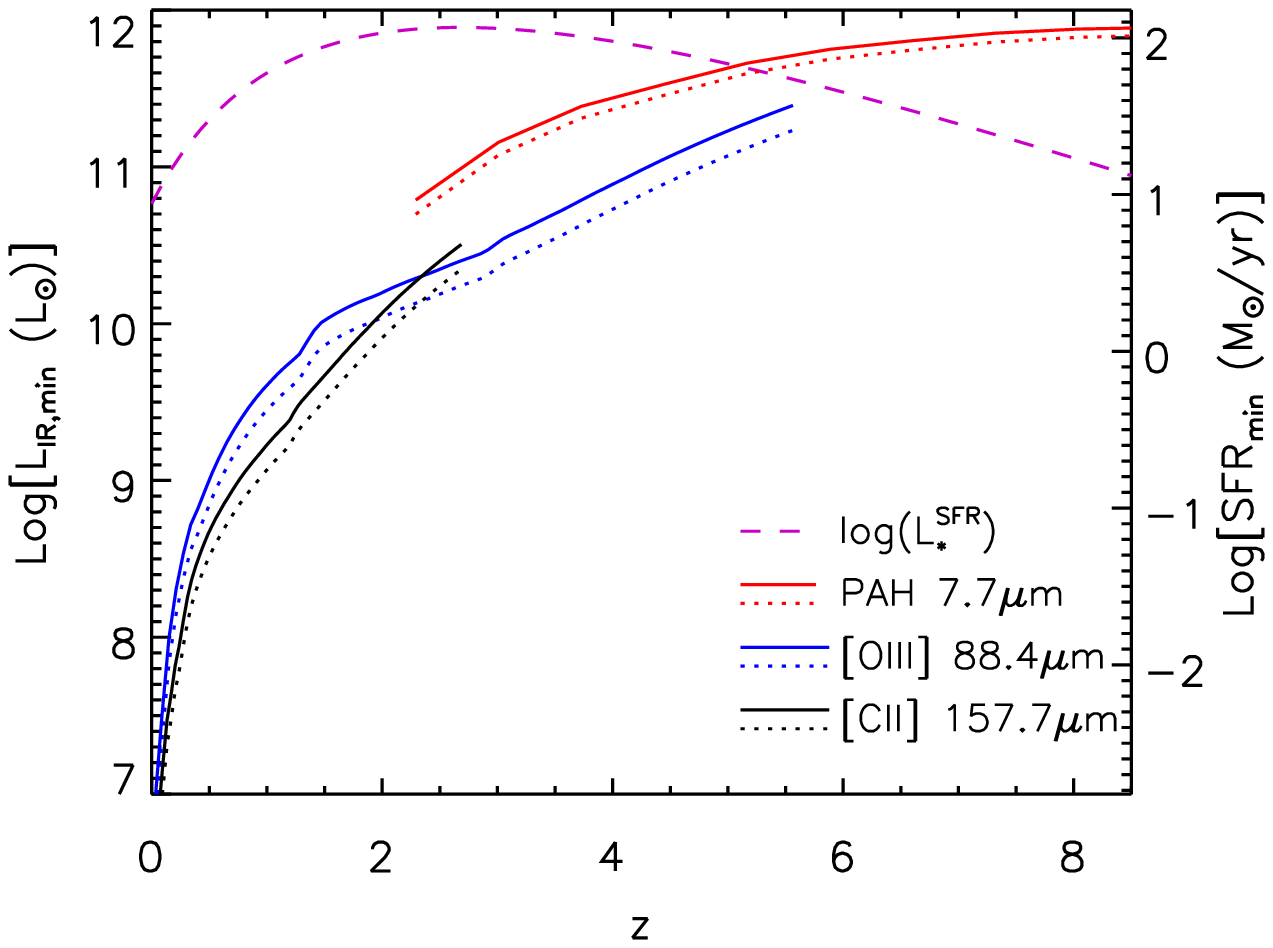}
		\includegraphics[trim=1.3cm 0.7cm 1.7cm 1.9cm,clip=true,width=0.49\textwidth,
    angle=0]{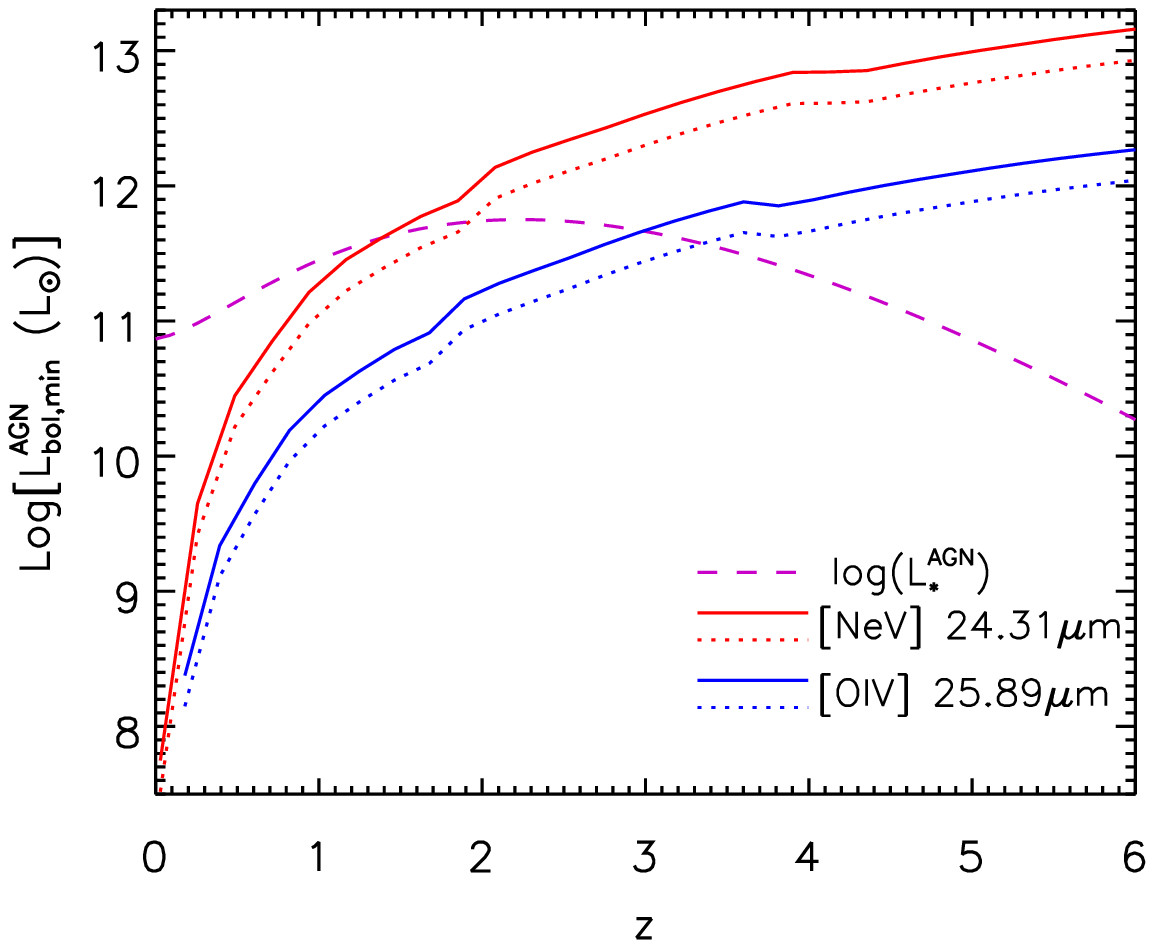}
			}
  \caption{Minimum IR luminosity - SFR (left panel, left and right y-axis respectively) and
  minimum AGN bolometric luminosity (right panel) detected,
  as a function of redshift, with a 5.9\,m (solid lines) and a 9.1\,m (dotted lines) OST survey
  with an areal coverage of 1\,deg$^{2}$, through spectroscopic detections of key SF and AGN lines.
  The dashed purple lines show, for comparison, estimates of the redshift-dependent characteristic
  luminosity, $L_\star$, of dusty galaxies and AGNs, respectively (see text).}
  \label{fig:SFR_AGN_Lbol}
\end{figure*}

As in \citet{Bonato2014b}, our simulations take into account both the SF and the
AGN components (see Sect.~\ref{sect:evol}). We calculated line luminosity functions
in 100 redshift bins with a width, $\Delta z$, of $\sim0.08$. For each redshift
bin, at the bin center, we extracted from the model SFR/IR luminosity functions
(for galaxies) or from the bolometric luminosity functions (for AGNs),
a number of objects, brighter than $\log(L_{\rm IR,
min}/L_{\odot})=8.0$, equal to that expected in $1\,\hbox{deg}^{2}$.

The luminosities of SF and AGN components of each source were randomly
extracted from the probability distributions that an object at redshift $z$ has
a starburst luminosity $L_{\rm IR, SF}$ or an AGN luminosity $L_{\rm AGN}$,
given the total luminosity $L_{\rm tot}=L_{\rm IR, SF}+L_{\rm AGN}$. The
probability distributions are given by \citet{Bonato2014b}.

The line luminosities associated with each component were randomly taken from
Gaussian distributions with the mean values and dispersions given in
Tables~\ref{tab:c_d_values} and \ref{tab:agn_a_b_values}. In this way we get,
at the same time, luminosities of {\it all} the lines of both the SF and
AGN components of each simulated object, allowing us to make interesting
predictions. For example, we can ask: For how many objects will both components be detectable
given the survey sensitivity? How many will be detectable in two or more lines?
Figure~\ref{fig:SED_protsph} shows examples of SEDs with AGN and SF components at $z=2$ and $z=5$.

We derived line luminosity functions by binning the simulated line
luminosities within each redshift bin. We repeated the simulations 1000 times
and averaged the derived line luminosity functions. Finally,
number counts were derived after assigning to each source a redshift drawn at random
from a uniform distribution within the redshift bin and computing the
corresponding flux. For more details, see \citet{Bonato2014b}.

\begin{figure}
\centering
		\includegraphics[trim=2.9cm 0.7cm 1.4cm 1.9cm,clip=true,width=0.49\textwidth,
    angle=0]{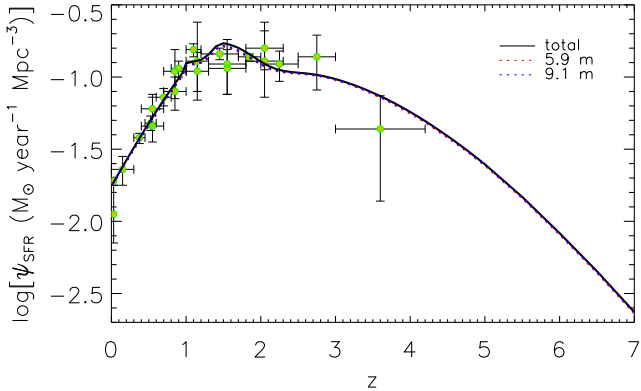}
  \caption{Cosmic SFR density resolved by an OST/OSS deep survey (1,000\,h over an area of 1\,deg$^{2}$) with
   a 5.9\,m (dotted red line) or a 9.1\,m (dotted blue dotted) telescope, as a function of z.
   The black solid line shows
  the total SFR density yielded by our model. The data points are from \citet{Madau14}.}
  \label{fig:SFRD}
\end{figure}

\section{Comparison between the 5.9\,m and 9.1\,m concepts}\label{sect:comparison_concepts}

The first question we want to address is whether, from the point of
view of extragalactic spectroscopic surveys with OST, there is strong scientific motivation to favor the more ambitious Concept~1 (9.1\,m telescope), compared to Concept~2 (5.9\,m telescope with a central obstruction). With
the performances mentioned in Sect.~\ref{sect:intro}, realistic surveys of
different depths could cover areas in the range 1--$10\,\hbox{deg}^2$.

\subsection{Number counts}

The survey depth at fixed observing time scales as the square root of
the ratio of the effective light collecting areas of the two concepts, i.e. as
$(52\,\hbox{m}^2/25\,\hbox{m}^2)^{1/2}=1.44$. To quantify the effect of such a
difference in depth, we have computed the number of 5\,$\sigma$ line detections
in 1000\,h by the OSS with both OST concepts, as a function
of the mapped area. The results are shown in Fig.~\ref{fig:detections}.

In all cases, the slopes of the counts are relatively flat at the
detection limits, implying that the number of detections in each line
increases more with the area than with the depth of the survey. We expect
$\sim 1.9 \times 10^{6}$ to $\sim 8.7 \times 10^{6}$ line detections for the
5.9\,m telescope, depending on the survey area, and $\sim 2.4 \times 10^{6}$ to $\sim 1.1 \times 10^{7}$ for
the 9.1\,m telescope. The number of detections in pure AGN lines ranges from $\sim
1.4 \times 10^{4}$ to $\sim 3.8 \times 10^{4}$ for the 5.9\,m telescope, depending on the survey area, and from
$\sim 2.1 \times 10^{4}$ to $\sim 6.0 \times 10^{4}$ for the 9.1\,m telescope.

We thus conclude that the decrease of the telescope size from 9.1\,m to
5.9\,m does not translate into a substantial worsening of the detection
statistics, at least in terms of counts.



\subsection{Redshift and luminosity distributions}

Of course, a comparison of the total number of detections is not enough
to answer the question posed at the beginning of this section, since going deeper may
allow us to reach poorly populated but interesting regions of the
redshift-luminosity plane.

Figure~\ref{fig:distributions_sources} shows the predicted redshift
distributions of sources detectable by OSS in at least one spectral
line over an area of 1\,deg$^{2}$ in 1,000\,h of observations for both OST
mission concepts. Although the larger telescope surpasses the smaller one in source detections, and does so
increasingly at greater redshifts, both
telescope sizes allow galaxy detections with good statistics up to
$z\simeq 8$, while the statistics are poor in both cases at higher redshifts.

The advantage of a larger telescope is somewhat greater if the objective is to
investigate galaxy-AGN co-evolution, which requires that sources be detectable
simultaneously in at least one SF line and one AGN line (Fig.~\ref{fig:distributions_sources_2components}). At
$z\ge 4.5$ the 9.1\,m telescope detects about twice as many sources as the
5.9\,m telescope, but the statistics are limited in both cases at this redshift.



To investigate the effect of telescope size on the
determination of the redshift-dependent IR luminosity functions (hence of the
SFR functions) and of AGN bolometric luminosity functions, we computed the minimum
$L_{\rm IR}$ and the minimum $L_{\rm bol}^{\rm AGN}$ corresponding to the
minimum luminosity of some bright lines detectable in a 1,000\,h survey of
$1\,\hbox{deg}^2$. To this end, we have used the appropriate value of the mean
line-to-IR or bolometric luminosity ratio. The results are shown, as a function
of redshift, in Fig.~\ref{fig:SFR_AGN_Lbol}.

The luminosities of the chosen SF lines (left panel) are proportional
to $L_{\rm IR}$, so the larger telescope would allow OST observers to reach luminosities lower
by a factor of 1.44. In the case of AGN lines, the ratio depends on luminosity
so the advantage of the larger telescope slightly increases with $z$, reaching
a factor of $\simeq 1.7$ at $z=6$ (right panel).

The right-hand scale on the left panel of Fig.~\ref{fig:SFR_AGN_Lbol}
shows the SFR corresponding to $L_{\rm IR}$, as given by eq.~(\ref{eq:KE}). Two
points should be noted here. First, as discussed in Sect.~\ref{sect:results},
in the case of the PAH lines eq.~(\ref{eq:KE}) breaks down at low
metallicities, so the correspondence between $L_{\rm IR}$ derived from
the PAH luminosity and the SFR is to be taken with caution, although the effect
of metallicity is expected to be minor at the relatively high values of $L_{\rm
IR}$ shown here. Second, eq.~(\ref{eq:KE}) holds for a Kroupa stellar
initial mass function (IMF). Recent ALMA measurements of multiple $CO$
transitions in four strongly lensed sub-mm galaxies at $z\simeq 2-3$
\citep{Zhang2018} have provided evidence for a top-heavy IMF, implying a greater
fraction of massive stars. Their preferred IMF would translate into a higher
$L_{\rm IR}/\hbox{SFR}$ ratio by about 40\%.

The dashed purple lines in Fig.~\ref{fig:SFR_AGN_Lbol} show, for
comparison, estimates of the characteristic luminosity, $L_\star$, as a
function of redshift, for dusty galaxies and AGNs (left- and right-hand panel,
respectively). The plotted values of $L_\star(z)$ refer to the modified
Schechter function defined by eq.~(9) of \citet{Aversa2015}. In the case of
dusty galaxies, $L_{\rm IR,min}<L_\star$  up to $z\simeq 5.9$ or $z\simeq 6.2$
for the 5.9\,m and the 9\,m telescope, respectively. This OST/OSS will then
provide a good description of the evolution of the bulk of the IR luminosity
function, hence of the IR luminosity density, up to these redshifts. In the AGN
case, $L_{\rm bol,min}<L_\star$  up to $z\simeq 3$ or $z\simeq 3.3$ for the two
telescope sizes.

Finally, Fig.~\ref{fig:SFRD} shows the fractions of the SFR density resolved by
the deep spectroscopic surveys as a function of z for
the two telescope sizes. The SFR density is almost fully resolved in both
cases ($\simeq 97\%$ with the 5.9\,m concept, $\simeq 98\%$ with the 9.1\,m
telescope).

\subsection{Confusion}\label{subsec:line_confusion}

\begin{figure*}
  \hspace{+0.0cm} \makebox[\textwidth][c]{
    \includegraphics[trim=4.4cm 6.3cm 4.9cm 2.6cm,clip=true,width=0.99\textwidth,
    angle=0]{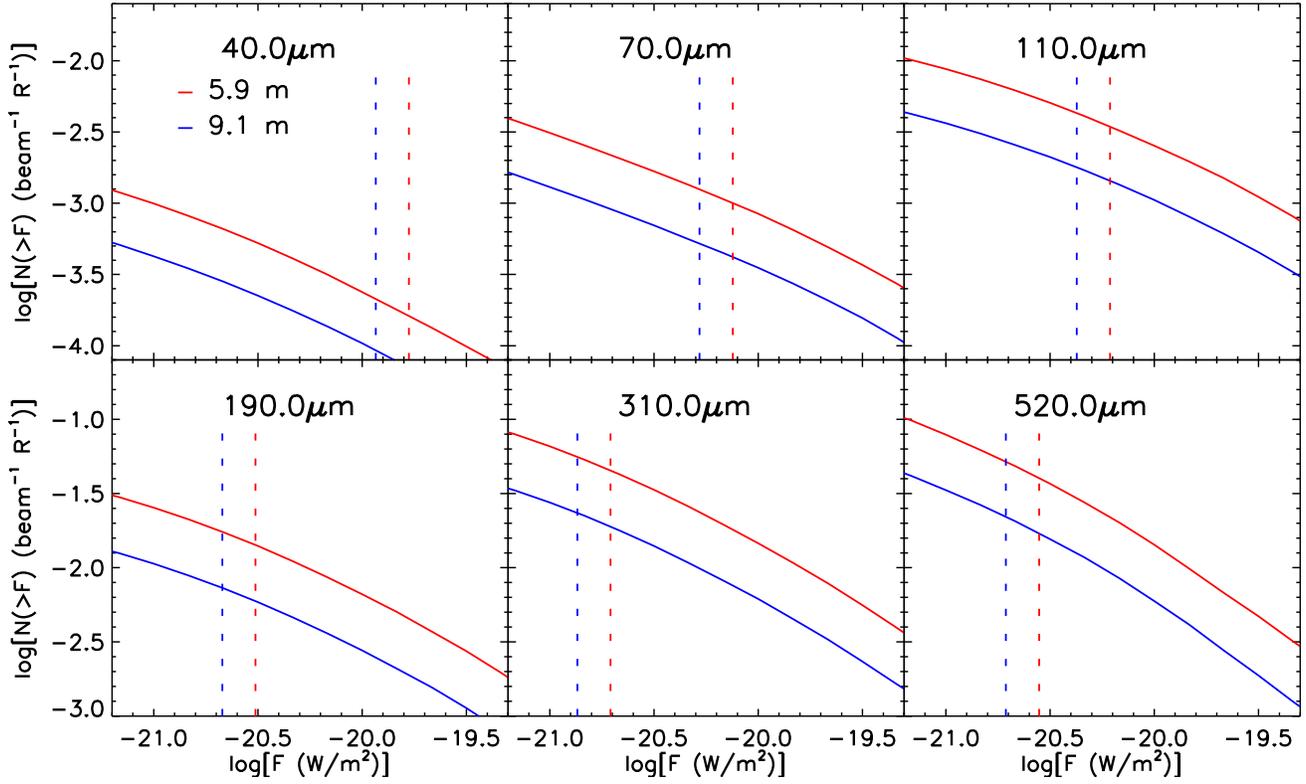}
  }
  \caption{Total integral line number counts per beam and per spectral resolution element
  at different wavelengths within the OST/OSS spectral coverage for the 5.9\,m (red) and
  the 9.1\,m (blue) telescopes. The vertical dashed lines represent the detection limits for
  surveys of 1,000\,h covering 1\,deg$^{2}$.}
  \label{fig:line_confusion}
\end{figure*}

The telescope in both OST concepts is diffraction limited at 30\,$\mu$m.
Throughout  most of the OSS wavelength range, a larger telescope provides
better angular resolution, and will be less prone to spatial confusion. Is
source confusion an issue, especially for the 5.9\,m telescope? Since
spectroscopy adds a third dimension, confusion is in general far less
significant than would be the case for a continuum survey. In spectroscopy,
confusion arises if lines from different sources along the line of sight can
show up by chance in the same spectral and spatial resolution element.

Figure~\ref{fig:line_confusion} shows the total integral number counts of lines
per beam\footnote{Each beam has $\theta_{\rm FWHM}$ of 1.13$\lambda$/D and a
solid angle of $\theta_{\rm FWHM}^{2}\times\pi/(4\ln2)$}  and per spectral
resolution element at different wavelengths within the OST/OSS spectral
coverage, for the 5.9\,m and 9.1\,m telescope concepts. The line detection limits for a
deep survey of 1\,deg$^{2}$ are represented by the vertical dashed lines.
Even in the worst case (longest wavelength and smaller telescope size), the
source density per resolution element is $<1/22$, i.e. the confusion is only
marginal, and observations at shorter wavelengths, where the detection limits
are well above the confusion limit, will help to resolve confusion in three (spatial-spectral) dimensions. Thus, line
confusion is never an issue for the 9.1\,m telescope and realistic exposure
times. The 5.9\,m telescope option can hit the confusion limits only at the
longest wavelengths, and only for very deep surveys ($\ge 1,000\,$h over
$1\,\hbox{deg}^2$).

\section{Surveys with the 5.9\,m OST}\label{sect:survey_strategy}

\begin{figure*}
  \hspace{+0.0cm} \makebox[\textwidth][c]{
    \includegraphics[trim=0.5cm 4.1cm 1.3cm 0.0cm,clip=true,width=0.99\textwidth,
    angle=0]{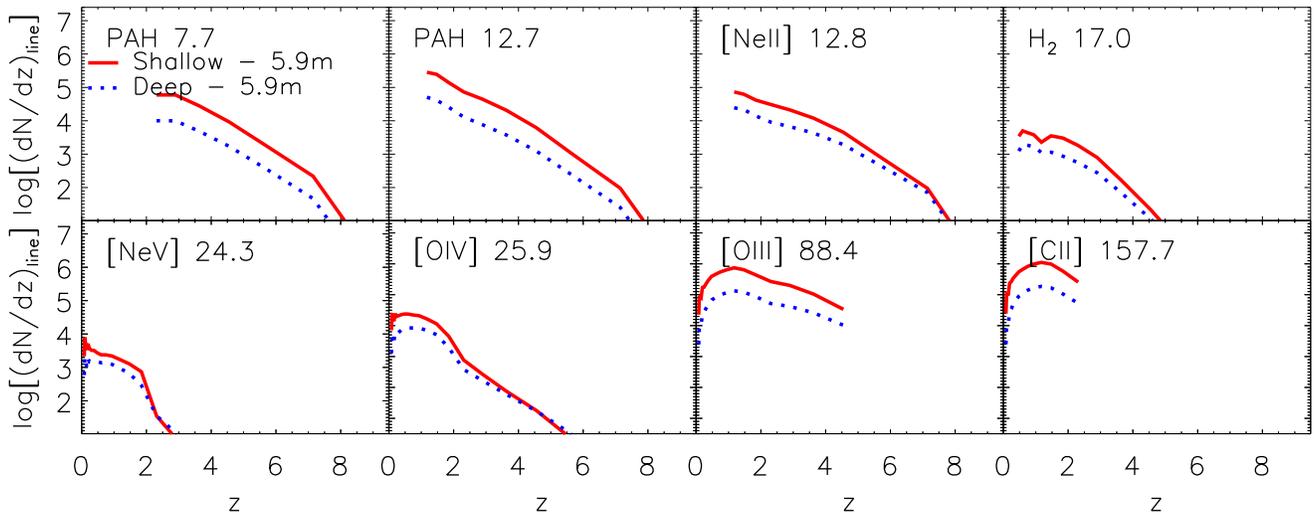}
  }
  \caption{Predicted redshift distributions of galaxies detected in the three brightest AGN
lines ([NeV]\-14.32, [NeV]\-24.31, and [OIV]\-25.89\,$\mu$m) and in selected lines indicating
star formation, namely those with relatively high detection rates. The predicted distributions pertain to shallow (solid red lines,
  10\,deg$^{2}$ areal coverage) and deep (dotted blue lines, 1\,deg$^{2}$ areal coverage) surveys,
  each conducted in 1000\,h of observing time with a 5.9\,m telescope.}
  \label{fig:nz}
\end{figure*}

\begin{figure*}
    \includegraphics[trim=3.3cm 3.6cm 4.1cm 0.2cm,clip=true,width=0.99\textwidth,
    angle=0]{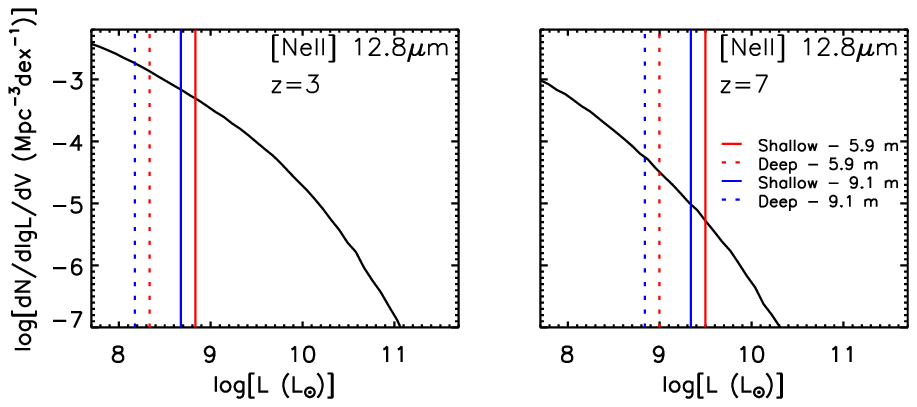}
\vskip-4cm
    \includegraphics[trim=3.3cm 5.8cm 4.1cm 0.2cm,clip=true,width=0.99\textwidth,
    angle=0]{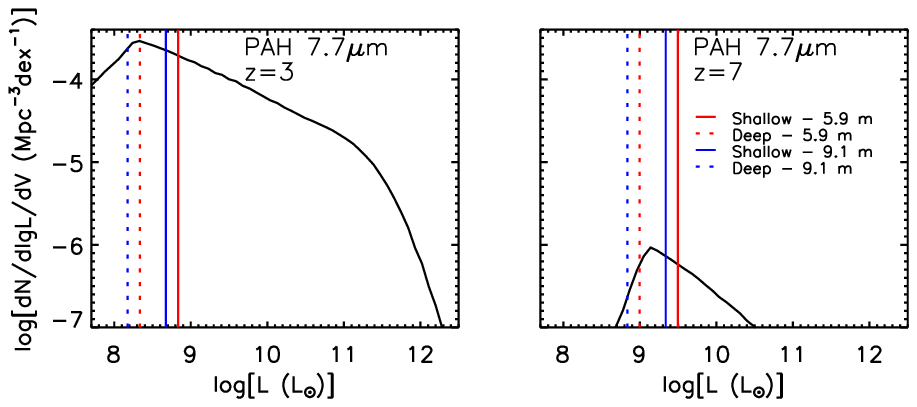}
  \caption{Luminosity functions of the fine structure line [NeII]\,$12.8\,\mu$m (upper panels) and the
  PAH\,$7.7\,\mu$m line at $z=3$ and $z=7$. The decline of the faint end of the PAH luminosity function is due
  to the effect of decreasing metallicity.}
  \label{fig:LF}
\end{figure*}

The analysis presented in the previous section has not exposed any
scientifically compelling advantage of the 9.1\,m telescope over the 5.9\,m
telescope. In this section we focus on the latter option and compare in more
detail the outcome of surveys covering different areas at fixed total observing
time, considering the cases of a wide and shallow ($10\,\hbox{deg}^2$ area) and
a deep ($1\,\hbox{deg}^2$) survey, each conducted in an observing time of
1,000\,h.

In Fig.~\ref{fig:nz} we display the redshift distributions of galaxies observed
in the three brightest AGN lines ([NeV]\-14.32, [NeV]\-24.31, and
[OIV]\-25.89\,$\mu$m) and in some of the most often detected star formation
indicators.  Obviously, only the shortest wavelength lines can be detected up
to very high redshifts. The figure clearly shows that the OST is so sensitive
that even the shallow survey can reach the highest redshifts. Again, the number
of detections increases more with increasing survey area than with increasing
depth. Hence, the shallow survey yields better statistics than the deep one,
except at the highest redshifts, and only for certain lines. Even in the most
favourable cases, the deep survey offers only minor advantages from a
statistical perspective.

A factor of 10 decrease in the surveyed area at fixed total observing
time translates into a decrease in the minimum
detectable line luminosity at given $z$ by a factor of $10^{1/2}\simeq 3.16$. In
the case of the fine structure lines, such as [NeII]\,$12.8\,\mu$m (upper
panels of Fig.~\ref{fig:LF}), which are estimators of the total (dust obscured
and unobscured) SFR, going deeper simply means extending the luminosity function to fainter levels. In the case of the
[NeII]\,$12.8\,\mu$m line, the ratios of numbers of detections \textit{per unit
area} between the deep and the shallow surveys are $\simeq 3.2$ and 7.5
at $z=3$ and $z=7$, respectively, for a 5.9\,m telescope. These ratios do not
compensate for the factor of 10 difference in survey area, so the total
number of detections is larger for the wide, shallow survey.

In the case of PAH lines the minimum detectable luminosity is in the
range where their emission is substantially reduced due to a decrease in
metallicity, as discussed in Sect.~\ref{sect:results}, and a corresponding dip is seen at the
low-luminosity end of the PAH luminosity function. Hence, the
increase in the number of PAH detections \textit{per unit area} with decreasing
detection limit is rather modest. For example, the ratio of the number of detections of the
$7.7\,\mu$m line between the deep and shallow surveys is $\simeq 1.7$ at
$z=3$ and $\simeq 2.1$ at $z=6$ (lower panels of Fig.~\ref{fig:LF}). The
increase in the total number of detections {per unit area} is also small since
PAHs are the dominant contributors. On the other hand,
this luminosity range carries information about the poorly known dependence of
$L_{\rm IR}$ on metallicity and on the impact of varying metallicity on the PAH
emission.

According to our model, the brightest IR AGN line, [OIV]$25.89\mu$m, will be
detected in $\sim33,000$ sources by the shallow survey and in $\sim12,000$
sources by the deep one, reaching, in both cases, $z \sim 5.5$. The shallow
survey will yield about 750,000 PAH detections up to $z \sim 8.5$, the deep
survey about 135,000 up to $z \sim 7.5$. The counter-intuitive
decrease in the maximum redshift with increasing survey depth, at fixed
observing time, is due to the increasing scarcity of bright PAH lines at high
$z$, a consequence of decreasing metallicity: large-area surveys are needed to find rare sources.

The [CII]\,157.70$\mu$m line, the strongest line emitted by the cool gas in
galaxies ($<$10$^{4}$\,K; see \citealt{CarilliWalter2013}), is detectable up to
$z\sim 3.2$ by the two surveys. We predict that the line will be seen in about
1.6 million galaxies by the shallow survey and in $\sim 280,000$ galaxies by
the deep survey. Moreover, the shallow and deep surveys will yield $\sim
12,300$ and $\sim 4,500$ detections, respectively, up to $z \sim 5$, in the
four warm H$_{2}$ lines, which are good diagnostics of the turbulent gas (upper
right-hand panel of Fig.~\ref{fig:nz}).

\begin{figure}
\centering
		\includegraphics[trim=3.4cm 1.2cm 1.6cm 0.6cm,clip=true,width=0.49\textwidth,
    angle=0]{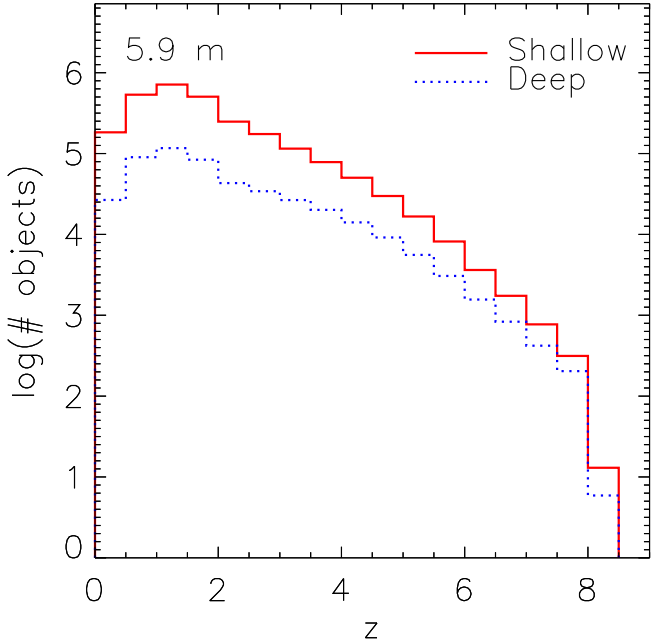}
  \caption{Predicted global redshift distributions of galaxies detected in spectral lines
  by the shallow (10\,deg$^{2}$) and deep (1\,deg$^{2}$) surveys with a 5.9\,m telescope.
  Each survey has a total amount of observing time equal to 1000\,h.}
  \label{fig:distributions_sources_5m}
\end{figure}

The conclusion that there is no need for exposures longer than those of
the shallow survey to reach the highest redshifts is confirmed by
Fig.~\ref{fig:distributions_sources_5m}, which illustrates the global redshift
distributions of sources detectable with the two notional surveys. The
shallow survey will detect $\sim2.7\times10^{6}$ galaxies, while the deep
survey will detect $\sim4.8\times10^{5}$. The peak of the distributions is at $z \sim 1.5$,
close to the peak of SF and BH accretion activity. The shallow and
deep surveys will detect about 31,000 and about 12,000 galaxies
at $z>5$, respectively.

\begin{figure}
\centering
		\includegraphics[trim=3.4cm 1.2cm 1.6cm 0.6cm,clip=true,width=0.49\textwidth,
    angle=0]{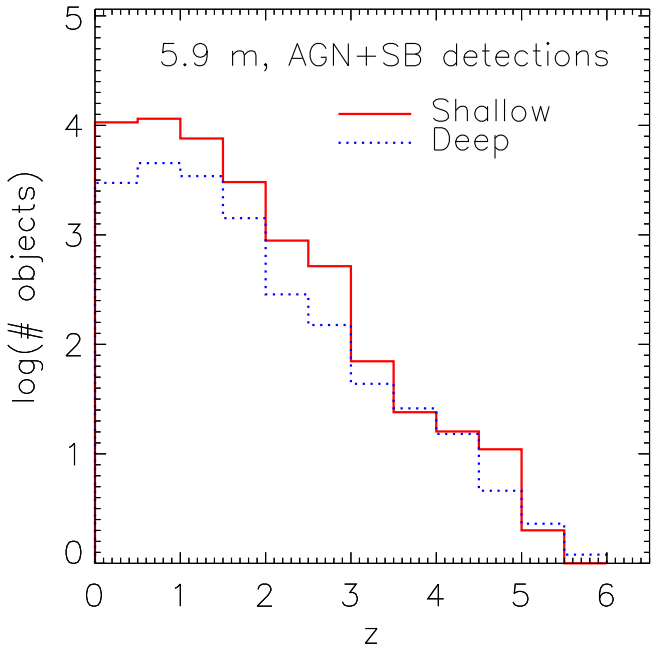}
  \caption{Predicted redshift distributions of galaxies
  detected simultaneously in at least one SF line and one AGN line by the deep (1\,deg$^{2}$)
  and shallow (10\,deg$^{2}$) surveys with a 5.9\,m telescope. Both surveys have a total amount of
  observing time equal to 1000\,h. As in the case of Fig.~\ref{fig:distributions_sources_2components},
  these distributions are almost identical to the redshift distributions of AGNs detected by the two surveys.}
  \label{fig:distributions_sources_2components_5m}
\end{figure}

For the fixed amount of observing time considered, the deep survey does not offer any significant advantage over the
shallow survey, even if we are looking for sources
detected simultaneously in at least one SF line and one AGN line.
Figure~\ref{fig:distributions_sources_2components_5m} shows that the deep
survey detects more such objects only at $z>5$, but even there the difference
is small and the statistics are poor. The total number of detections of
such high-$z$ sources are of $\sim3.4\times10^{4}$ and $\sim1.3\times10^{4}$ for the
shallow and deep surveys, respectively.

\begin{figure*}
  \centering
		\includegraphics[trim=3.4cm 1.2cm 1.6cm 0.6cm,clip=true,width=0.49\textwidth,
    angle=0]{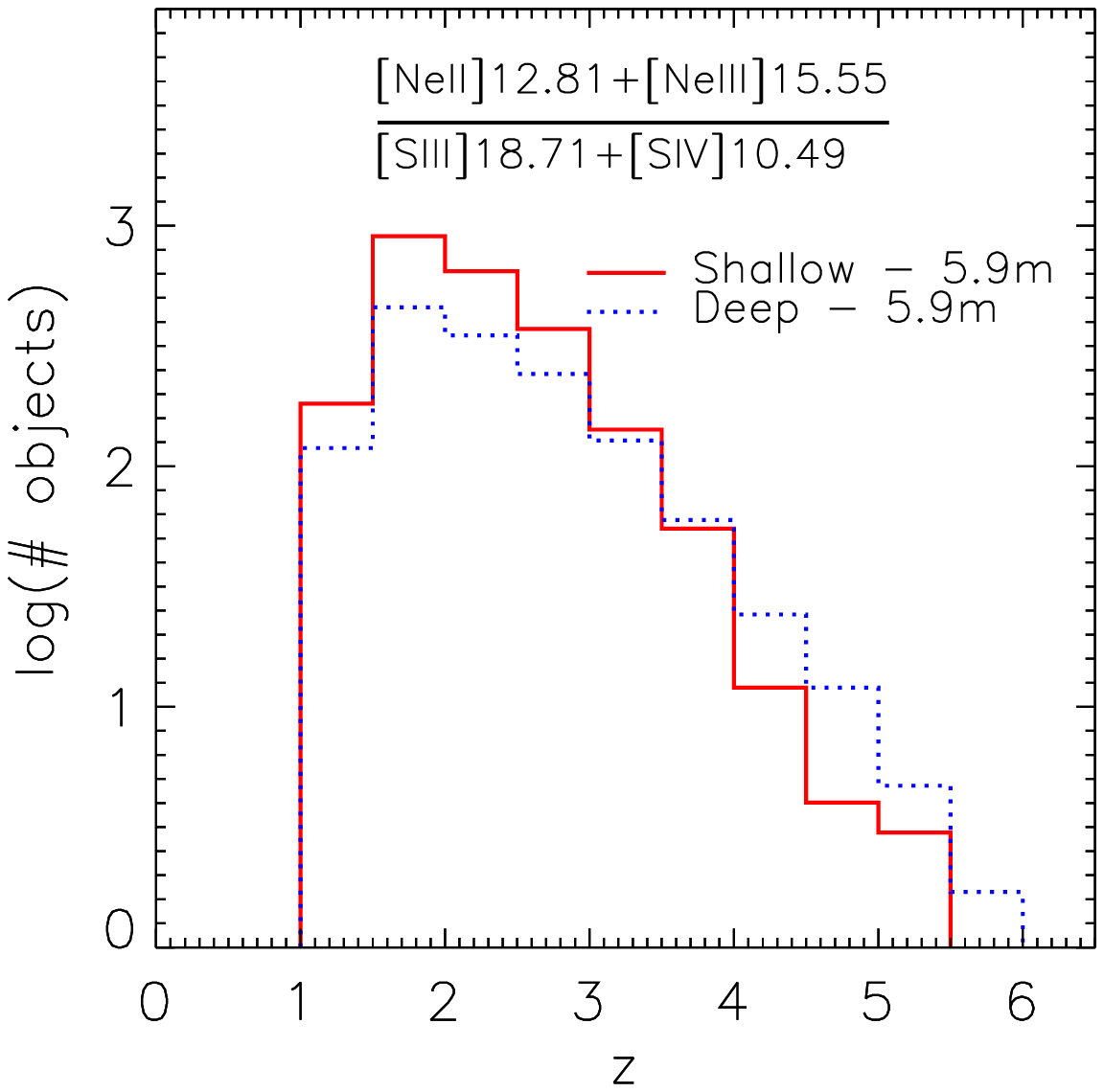}
		\includegraphics[trim=3.4cm 1.2cm 1.6cm 0.6cm,clip=true,width=0.49\textwidth,
    angle=0]{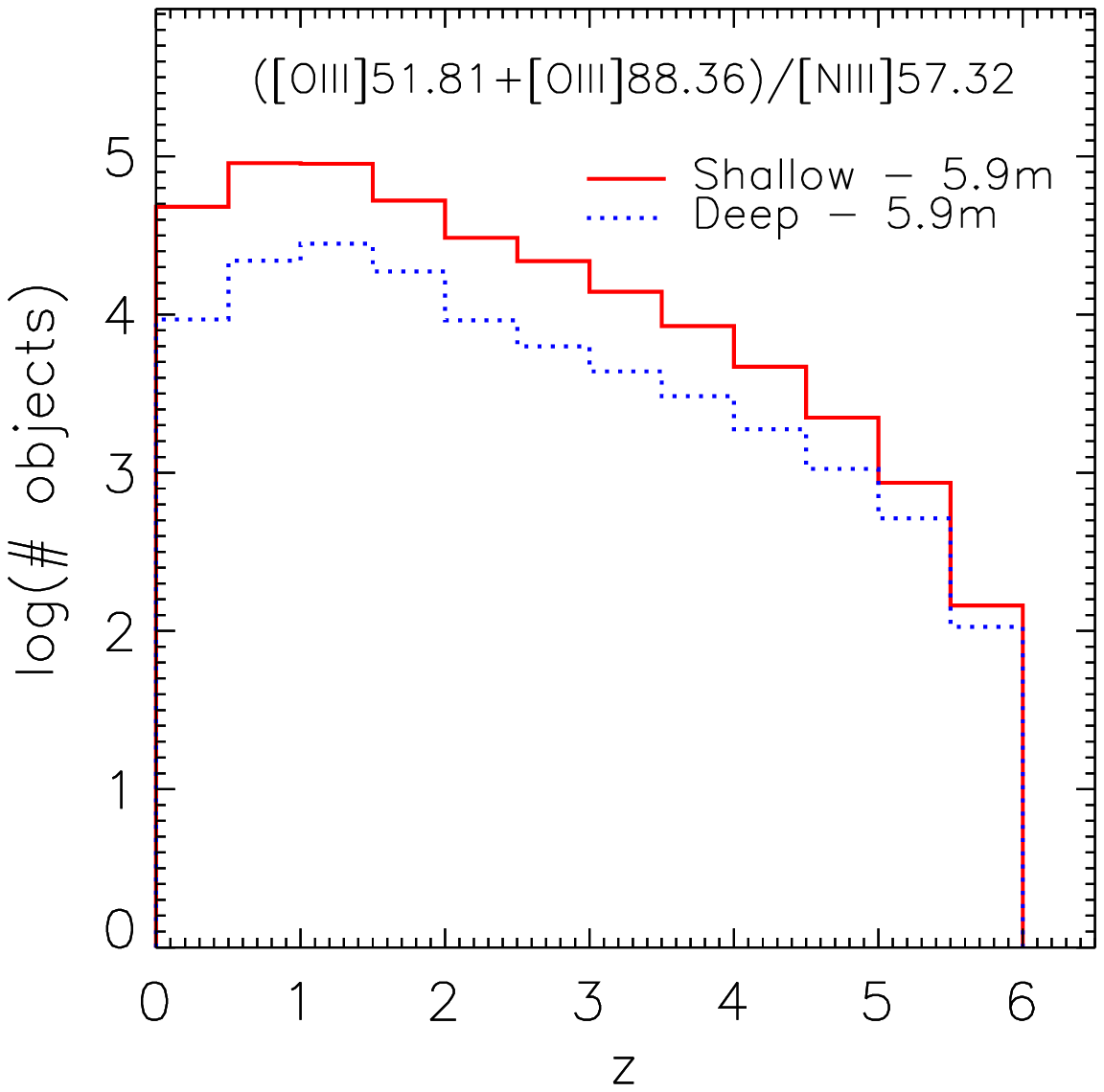}
  \caption{Left panel: predicted redshift distributions of
  galaxies detected simultaneously in the [NeII]\,12.81, [NeIII]\,15.55, [SIII]\,18.71, and
  [SIV]\,10.49$\mu\hbox{m}$ lines by the shallow (10\,deg$^{2}$) and deep (1\,deg$^{2}$)
  surveys with a 5.9\,m telescope. Both surveys have a total amount of observing time equal to 1000\,h.
  Right panel: the same, but for the simultaneous detection of the [OIII]\,51.81, [OIII]\,88.36, and
  [NIII]\,57.32$\mu\hbox{m}$ lines.}
  \label{fig:distributions_metallicity_5m}
\end{figure*}

Next consider the two metallicity diagnostics described in
Sect.~\ref{sect:intro}, involving the following spectral lines: 1)
[NeII]\,12.81, [NeIII]\,15.55, [SIII]\,18.71, and [SIV]\,10.49$\mu\hbox{m}$; or 2)
[OIII]\,51.81, [OIII]\,88.36, and [NIII]\,57.32$\mu\hbox{m}$.
Figure~\ref{fig:distributions_metallicity_5m} shows the redshift distributions
of sources detectable by the shallow and deep surveys conducted with a 5.9\,m
telescope in 1000\,h of observing time, for which these two metallicity
diagnostics can be built thanks to the simultaneous detection of the two sets
of lines.

The deep survey is advantageous for the first metallicity diagnostic at
$z>4$, although the total number of detections is larger for the shallow survey
($\sim 2,300$ vs  $\sim 1,400$). In the case of the second metallicity
diagnostic the shallow survey is better at all redshifts: the total number of
detections is $\simeq 3.6 \times 10^{5}$ with $\sim 1,000$ galaxies at $z>5$,
versus $\simeq 1.0\times 10^{5}$ with $\sim 600$ galaxies at $z>5$ for the
deep survey.

\section{Discussion}\label{sect:discussion}

\subsection{Galaxy-AGN co-evolution}

As shown in Fig.~\ref{fig:distributions_sources_2components}, we predict the
simultaneous detection of SF and AGN lines for a large sample (a few tens of
thousands) of galaxies. The brightest IR AGN line - [OIV]$25.89\mu$m - will
allow us to estimate the bolometric luminosity of the AGN component of the
sources up to $z \sim 5.5-6$, while starburst lines provide estimates of the IR
luminosity, and hence the SFR up to $z \sim 8-8.5$ (Fig.~\ref{fig:nz}). In other
words, for the first time OST will enable observers to explore the physical
processes driving galaxy-AGN (co-)evolution throughout most of cosmic history,
with excellent statistics.

\begin{figure*}
\hspace{+0.0cm} \makebox[\textwidth][c]{
\includegraphics[trim=2.0cm 7.3cm 1.2cm 2.2cm,clip=true,width=0.99\textwidth,
    angle=0]{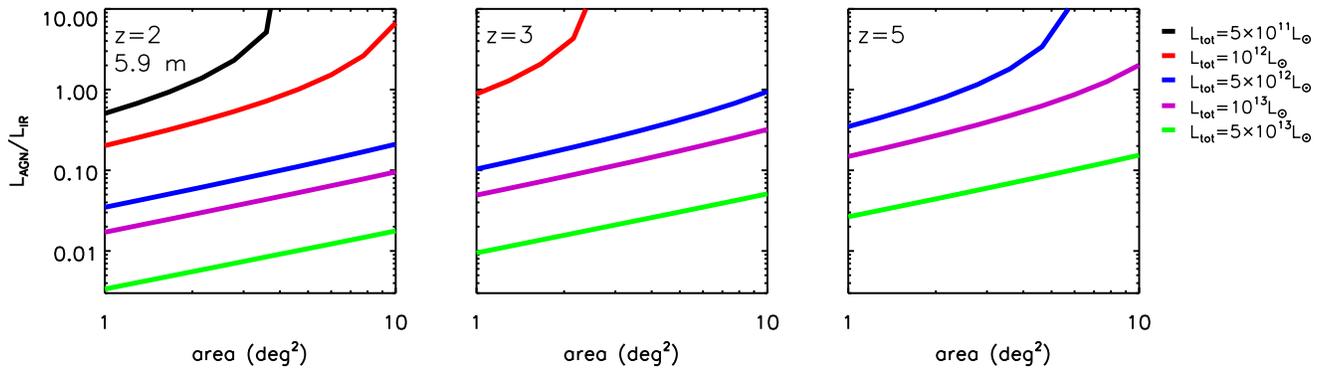}
  }
  \caption{Minimum ratios of AGN bolometric luminosities to SF IR luminosity $L_{\rm IR}\equiv L_{\rm IR, SF}$
  for which lines associated with both components (the [OIV]\,25.89$\mu$m and the [OIII]\,88.36$\mu$m lines, respectively)
  are detectable with the OST/OSS at $z=2$ (left), $z=3$ (centre), and $z=5$ (right), as a function of the mapped area for five values of $L_{\rm tot}=L_{\rm AGN}+L_{\rm IR, SF}$. }
  \label{fig:AGN_SF_ratio}
\end{figure*}

Figure\,\ref{fig:AGN_SF_ratio} illustrates the extent to which galaxy-AGN
co-evolution can be tested with OSS spectroscopy at three redshifts
($z= 2$, 3 and 5) around and beyond the peak of SF and BH accretion activity.
The figure shows, as a function of area mapped, the minimum ratio of
the AGN bolometric luminosity, $L_{\rm AGN}$, to the IR luminosity, $L_{\rm IR}
\equiv L_{\rm IR, SF}$, of the host galaxy for which OST can detect
both the [OIV]\,25.89$\mu$m and the [OIII]\,88.36$\mu$m lines. As shown
above, these lines are good indicators of the AGN and SF luminosities, respectively.
The curves refer to five values of the total bolometric luminosity, $L_{\rm
tot}=L_{\rm AGN}+L_{\rm IR, SF}$, of the source\footnote{During the
dust-enshrouded active SF phase, $L_{\rm IR, SF}$ is approximately the
bolometric luminosity of the galaxy.}, ranging from $5\times10^{11}$ to
$5\times 10^{13}\,L_\odot$\footnote{Note that the density of these
(ultra-)luminous IR galaxies quickly decreases with increasing redshift.}.

During the dust-obscured intense SF phase, the BH is expected to accrete at, or
perhaps slightly above, the Eddington limit. At the end of this phase, its
mass should, on average, be close to that given by the correlations with host
galaxy properties. The corresponding AGN bolometric luminosity, $L_{\rm AGN}$,
can be about an order of magnitude greater than the IR luminosity of the host galaxy,
$L_{\rm IR,SF}$. For a source with $L_{\rm tot}=10^{13}L_{\odot}$, the
deep survey (1\,deg$^{2}$) could detect both AGN and SF lines down to $L_{\rm
AGN}/L_{\rm IR, SF}\simeq 0.017$ at z$=$2, $\simeq 0.05$ at $z=3$ and $\simeq
0.15$ at $z=5$.

For Eddington-limited accretion, the corresponding BH masses are $\simeq
5.2\times 10^6\,M_\odot$, $\simeq 1.5\times 10^7\,M_\odot$, and $\simeq
4.1\times 10^7\,M_\odot$ at redshifts 2, 3, and 5, respectively. The final total
luminosity of these objects is AGN dominated; using the \citet{Cai13} Eddington
ratios, we find that the final BH masses are $3.1\times 10^8\,M_\odot$,
$2.6\times 10^8\,M_\odot$ and $1.4\times 10^8\,M_\odot$, respectively. Thus the
OST/OSS surveys can detect AGNs when their BH masses are factors $\sim 60$ (at
$z\simeq 2$), $\sim 18$ (at $z\simeq 3$), and $\sim 3$ (at $z\simeq 5$) lower
than the final values, for a galaxy with a total IR luminosity of
$10^{13}\,L_\odot$. The probed BH mass growth range will be larger for the rare
more luminous sources. Thus, the [OIV]$25.89\mu$m line, which is not much
affected by dust obscuration, is an efficient tool for taking a census of
faint AGNs inside dusty star-forming galaxies.

Existing information on co-evolution is very limited and based on a combination
of X-ray and far-IR/sub-mm data. X-ray observations are currently the most
efficient way to reliably identify AGNs. However, X-ray emission associated
with stellar processes can reach luminosities of $\simeq
10^{42}\,\hbox{erg}\,\hbox{s}^{-1}$ \citep[cf., e.g.,][]{Lehmer2016}, and this
makes the identification of lower luminosity AGNs challenging or impossible.
OST observations will not have this limitation;  OST will be able to detect
less luminous AGNs in the local Universe.

For example, using an X-ray bolometric correction of 20, as appropriate at the
luminosities considered \citep{Lusso2012}, we find that an OST
(5.9\,m) deep survey with OSS can detect AGN line signatures via
the [OIV]$25.89\mu$m line up to a factor of $\sim$12 fainter at $z \simeq
0.2$. The redshift distribution of galaxies detected simultaneously in
at least one SF line and one AGN line
(Fig.~\ref{fig:distributions_sources_2components_5m}) implies that the OST
will detect the AGN component of about 3,500 galaxies at $z\le 0.2$.

Alternative methods to detect faint AGNs include sub-arcsec mid-IR imaging to
isolate the AGN component from the emission due to SF \citep[currently possible
only for nearby AGNs; e.g.,][]{Asmus2015}, and mid-IR photometry in the
wavelength range 3--$30\,\mu$m \citep[e.g.,][]{Assef2013}. The latter approach
exploits the fact that the AGN emission is ``hotter'' than that due to
star formation to decompose the observed spectral energy distribution,
separating the two components; it works primarily for intrinsically luminous
but heavily obscured AGNs.

Current studies of the connection between AGN activity and SF have been carried
out using either the X-ray emission of star-forming galaxies
\citep[e.g.,][]{Delvecchio2015} or the SF properties of X-ray selected AGNs
\citep[e.g.,][]{Stanley2015}. In both cases, caveats apply to the
interpretation of the results. X-ray detections were found in only a small fraction of the
star-forming galaxies selected from \textit{Herschel} surveys. For example, only about 10\% of galaxies in the
\citet{Delvecchio2015} sample were detected in X-rays. For the overwhelming
majority of galaxies, only average AGN accretion rates (or bolometric
luminosities) could be determined by stacking. The X-ray selected sample by
\citet{Stanley2015} has the opposite problem: only upper limits to the IR
luminosity due to SF could be placed for 75.4\% of the sources. In both
cases, a large fraction of objects have only photometric redshift estimates.
Spectroscopic redshifts are available for 53\% of the
\citet{Stanley2015} sample, while for the one by \citet{Delvecchio2015} the
fraction of spectroscopic redshifts is $\simeq 60$\% at $z\le 1.5$, but
drops to 20\% at $z\simeq 2$. For comparison, the proposed OST/OSS surveys
will detect both SF and AGN spectral signatures for tens of thousands of
sources up to $z \sim 5.5-6$.

\subsection{Evolution of gas metallicity}

Detection of the bright PAH bands up to z$\sim$8.5 will allow us, for the first
time, to follow their entire history up to the re-ionization epoch, and
therefore to explore the properties of the ISM and to address many open
questions regarding their formation, destruction and dependence on metallicity.

Most galaxies will be detected by the OST/OSS in at least 2 lines ($\sim$60\%
of the galaxies observed with the 5.9\,m OST shallow survey, $\sim$70\% with
the deep survey). Therefore, we will have robust spectroscopic
redshift measurements. 
Figure\,\ref{fig:distributions_metallicity_5m} shows that the OST will
allow us to  measure metallicity in galaxies up to z$\sim$6.

The results presented in this paper pertain to blind surveys.  Obviously, the
OST/OSS could go deeper by targeting known high-$z$ galaxies, but blind,
unbiased surveys are needed to obtain unbiased censuses of the SFR density and the BH accretion rate density.

\begin{figure}
\centering
		\includegraphics[trim=1.5cm 0.7cm 3.3cm 1.9cm,clip=true,width=0.49\textwidth,
    angle=0]{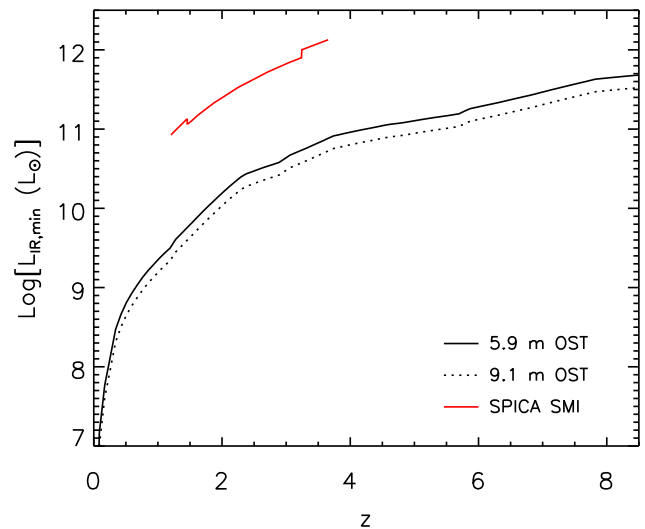}
  \caption{Comparison of the minimum $L_{\rm IR}$ reachable with SPICA/SMI and OST/OSS,
  as a function of redshift. The upper solid (red) line shows the estimate
  by \citet[][lower panel of their Fig.~4]{Kaneda2017}
  for galaxies with 100\% SF detected via the 6.6, 7.7, 8.6 and $11.3\,\mu$m PAH
  features in their ``deep survey'' (565.5\,h of observing time, without overheads,
  over an area of $1\,\hbox{deg}^2$). The lower solid (black) line and the dotted line show
  the corresponding minimum $L_{\rm IR}$ for OST/OSS
  surveys with a 5.9\,m or a 9.1\,m telescope, respectively,
  over the same area and in the same observing time. The difference
  between the SPICA/SMI and the OST/OSS limiting $L_{\rm IR}$ is actually bigger than shown in this figure
  because the OST/OSS estimate takes into account the decrease of PAH luminosities
  with decreasing gas metallicity, while the SPICA/SMI estimate does not. }
  \label{fig:SMI}
\end{figure}

\subsection{Caveats}\label{sect:caveats}

Although our reference model was validated against a wide variety of
data, it necessarily relies on a number of simplifying assumptions and, as any
other model, cannot match the enormous variety of individual galaxy and AGN
properties.

Recent observations \citep{Dowell2014, Asboth2016, Ivison2016} have
yielded estimates of the space density of distant ($z\simgt 4$) dusty star-forming
galaxies well in excess of model predictions. As shown in the Appendix,
the model matches quite well the redshift distribution of such sources
estimated by \citet{Ivison2016}, but underestimates by a factor of $\simeq 2.5$
their (highly uncertain) total counts.

The interpretation and even the reality of the excess are still unclear
 \citep{Bethermin2017, Donevski2018}. If real, the excess would imply that the
 OST will detect several times more high-$z$ galaxies. Note
 that most redshift estimates used to search for high-$z$ galaxies are
 photometric, based on observations in a few sub-mm bands, and as such are quite
 uncertain. The OST/OSS will provide fully spectroscopic redshift distributions.
 These powerful tools can be used to discriminate among models that perform similarly well
 on higher-level statistics, such as source counts \citep[see, e.g., Fig.~9 of][]{Ivison2016}.

Recent observations of the four strongest CO emitters at $z\simeq 2-3$ selected
 from the literature \citep{Zhang2018} have provided evidence of an IMF with a higher
fraction of massive stars than the Chabrier IMF adopted for our reference model.
To what extent would our predictions be affected if this is a general property
of high-$z$ star-forming galaxies? The answer is: the effect would be only
marginal. This is because our line luminosity functions are derived from IR
luminosity functions, which are observationally determined up to $z\simeq 4$
\citep{Gruppion2013} and are well reproduced by the model. If we change the IMF,
we must also modify other parameters to preserve agreement with the
data.

Further uncertainties are associated with the relationships between
line and IR or bolometric luminosity. As pointed out in Sect.~\ref{sect:results},
in several cases there are not enough data to accurately determine the slope of the
relation. Data themselves must be taken with caution since, although we have done the
selection as carefully as we could, it is possible that some line luminosities
attributed to star formation are contaminated by AGN contributions and vice-versa.
In Sect.~\ref{sect:results}, we also noted that data on high-$z$ galaxies
frequently refer to strongly lensed galaxies and are therefore prone to errors
in the corrections applied for gravitational magnification and differential lensing.

In consideration of the aforementioned caveats, the present results should be taken with
a grain of salt. They are a sort of best guess based on state-of-the art understanding.

\subsection{Comparison to other instruments}\label{sect:compar}

From Fig.\,\ref{fig:detections} we can see that 1000\,h of observing time with
both OST concepts will yield from a few to several tens of thousands of AGN
detections, and from a few to $\simeq 10$ millions of SF detections, depending
on the extent of the surveyed area. The detected sources will have
broad redshift distributions, extending beyond $z=8$ in the case of SF galaxies
(Figs.~\ref{fig:distributions_sources} and \ref{fig:distributions_sources_5m})
and up to $z\simeq 5.5$ in the case of AGNs
(Figs.~\ref{fig:distributions_sources_2components} and
\ref{fig:distributions_sources_2components_5m}).

Blind spectroscopic surveys of similar areas (1--$10\,\hbox{deg}^2$)
with the low-resolution Mid-infrared Instrument (SMI) on board the SPace IR
telescope for Cosmology and Astrophysics (SPICA) have been discussed by
\citet{Bonato2015} and, more recently, by \citet{Kaneda2017}. These surveys
will exploit the large field of view (12\,arcmin$\times$10\,arcmin), high
speed, low-resolution ($R\sim 50-120$) spectroscopic  and photometric camera,
covering the wavelength range $17-36\,\mu$m, and are focussed on the detection
of PAH features.

According to \citet{Kaneda2017}, the planned ``wide'' survey (covering
$10\,\hbox{deg}^2$ with an observing time of 435\,h, excluding overheads) will
detect $\simeq 81,600$ galaxies, while the ``deep'' survey ($1\,\hbox{deg}^2$
with an observing time of 565.5\,h, excluding overheads) will detect $\simeq
22,200$ galaxies. Both surveys will attain redshifts somewhat above $z=4$,
bracketing the peak of the cosmic SFR density.

We note that the numbers of SPICA/SMI detections are probably
overestimated because they do not take into account the decrease of the PAH
luminosity with decreasing metallicity [eq.~(\ref{eq:PAH})]. Since our
calculations take this effect into account, the comparison with the estimates
by \citet{Kaneda2017} is not straightforward. Figure~\ref{fig:SMI} shows that,
ignoring this difference, the OST/OSS goes more than one order of magnitude
deeper than the SPICA/SMI. Also the OST/OSS can reach $L_{\rm IR}$ below
$10^{12}\,L_\odot$ (i.e. SFRs below $\simeq 100 M_\odot\,\hbox{yr}^{-1}$) up to
$z\simeq 8$ while the SPICA/SMI is limited to $z\simlt 4$.

The SPICA Far Infrared Instrument \citep[SAFARI;][]{Roelfsema2018}
includes a grating spectrograph with low ($R = 300$) and medium ($R \sim
2,000-11,000$) resolution observing modes covering the $34-230\,\mu$m
wavelength range. However, it has a rather narrow FoV and was not designed as a
survey instrument. It is expected to detect, in 1000\,h of observing time, from
some to a few tens of thousands of galaxies and, like the SMI, it will be limited to
$z<5$ (\citealt{Bonato2014a, Bonato2014b, Gruppioni2016}). For comparison, the OST/OSS
surveys with a 5.9\,m telescope will detect a few tens of thousands of
galaxies at $z\geq 5$.

Ground-based (sub-)mm telescopes, such as the Atacama Large
(sub-)Millimeter Array (ALMA), the NOrthern Extended Millimeter Array (NOEMA),
and the Large Millimeter Telescope (LMT) are allowing us to extend the detection of
long-wavelength far-IR and sub-mm lines to higher redshifts.
Further substantial progress is expected with new instruments, such as
CCAT-prime \citep{Stacey2018} and the Atacama Large-Aperture Submm/mm
Telescope \citep[AtLAST;][]{Bertoldi2018}.


\section{Summary}\label{sect:summary}

The unprecedented sensitivity of the Origins Space Telescope with the OSS instrument will allow us to address a variety
of open  issues concerning the evolution of galaxies and AGNs. To illustrate its
potential we estimated the counts of 53 IR spectral lines (9 of them are
typical AGN lines). This was done by coupling relationships between line and IR
continuum luminosity (or bolometric luminosity in the case of AGNs) with the SFR
and luminosity functions by \citet{Cai13}, which accurately match the
observational data. We have adopted the relationships found by
\citet{Bonato2014a, Bonato2014b, Bonato2015, Bonato2017} for the SF
contribution to the emission of 22 IR fine-structure lines and 6 PAH lines, and
for the AGN emission of 27 IR lines. For the SF contribution of 3 lines, the
relationships were recalibrated using new data, and new relationships were
determined for 10 CO lines, for the SF contribution to 3 lines, and for the AGN
contribution to 5 lines. \textit{We took into account the dependence of the
starburst lines on gas metallicity.}

We have compared the outcomes of blind surveys with two concepts for the
OST, a 5.9\,m telescope (Concept 2), and 9.1\,m telescope (Concept 1). Our analysis found no
compelling scientific case for the larger telescope.

We then investigated in more detail the performance of the 5.9\,m
telescope option for different surveyed areas, at fixed observing time. The
number of line (and source) detections steadily increases with increasing area.
Focussing on two extreme cases -- surveys of 1 and $10\,\hbox{deg}^2$, each with
an observing time of 1,000\,h -- we find that both can detect emission lines
associated with star formation (allowing the estimate of the SFR) up to $z\simeq
8.5$ (i.e., all the way through the reionization epoch), and AGN lines (allowing
an estimate of the bolometric luminosity) up to $z\simeq 5.5$.

The larger area survey will detect $\sim 8.7 \times 10^{6}$ lines from $\sim
2.7 \times 10^{6}$ star-forming galaxies and $\sim 3.8 \times 10^{4}$ lines
from $3.5 \times 10^{4}$ AGNs. It will generally provide better statistics,
except at the highest redshifts where, however, the advantage of the deeper
survey is modest and the statistics are poor anyway. The deeper survey will do
significantly better only for some metallicity diagnostics at $z>4$, and to
investigate the poorly known dependence of the PAH emission (and, more
generally, of $L_{\rm IR}$) on metallicity.

Both surveys will almost completely resolve the cosmic SFR density. The
OST with OSS will provide estimates of the IR luminosity function down to below
the characteristic luminosity $L_\star$ up to $z\simeq 6$. Thus it will allow
us to reconstruct essentially the full SF history, complementing the
information from ultra-deep surveys at near-IR wavelengths, which miss heavily
dust-enshrouded galaxies. It will also allow us to learn about the evolution of
metallicity and the physical conditions in the interstellar medium up to very
high redshifts. Note the key role played by the large OST/OSS field of view: we
are dealing here with rare sources that cannot be picked up in statistically
significant numbers by surveys with the James Webb Space Telescope, which can
hardly be expected to carry out deep surveys over areas much larger than
$100\,\hbox{arcmin}^2$ \citep[e.g.,][]{Finkelstein2015}.

The model also gave us the probability distributions that an object at
redshift $z$  has a starburst IR luminosity $L_{\rm IR,SF}$ or an AGN
bolometric luminosity $L_{\rm AGN}$, given the total luminosity $L_{\rm tot}=
L_{\rm IR,SF} + L_{\rm AGN}$. Our approach to deriving the line luminosity
functions allowed us to obtain, at the same time, luminosities of all the
lines of both the SF and the AGN components of each simulated object. We could
thus estimate for how many objects both components will be detectable, given the
survey sensitivity and the number of objects that will be detectable in two or more lines.

This allowed us to investigate the OST/OSS potential to provide
insight into galaxy-AGN co-evolution. We found that the OST will be able to probe
BH mass growth by large factors, especially, but not only, at moderate
redshifts ($z\lesssim 3$) and for luminous sources (star-formation plus AGN
luminosity $\gtrsim 10^{13}\,L_\odot$). Spectroscopic surveys with OST will thus be a powerful tool to get
a census of faint AGNs inside dusty star-forming galaxies over a broad redshift
range.

Compared to SPICA/SMI, which is also expected to carry out blind surveys
similar to those discussed here, the OST/OSS will reach, in the same amount of observing
time, fluxes more than a factor of 10 fainter and will detect galaxies with
$L_{\rm IR}$ below $10^{12}\,L_\odot$ up to $z\simeq 8$, while the SPICA/SMI is
limited to $z\lesssim 4$.

On the basis of the existing simulations (\citealt{Bonaldi2018}) and  our
estimates, we expect that the OST/OSS will be able to measure redshifts of a
substantial fraction of the galaxies detected by the Square Kilometre Array
(SKA\footnote{\url{www.skatelescope.org}}) deep surveys. The SKA will provide
an independent estimate of the SFR in these galaxies, and will also not be
affected by obscuration (see \citealt{Mancuso2015}, e.g. their Fig.~12), but
SKA measurements will be susceptible to contamination by nuclear radio
activity. Together, OST and SKA will provide a more comprehensive view of the
galaxy/AGN evolutionary scenario over a broad frequency range.



\begin{acknowledgements}
We thank Roberto Maiolino and Alessandro Bressan for enlightening
suggestions, and the anonymous referee for useful comments that led to
substantial improvements in this paper. The study acknowledges partial financial
support by the Italian Ministero dell'Istruzione, Universit\`{a} e Ricerca
through the grant `Progetti Premiali 2012 - iALMA' (CUP C52I13000140001). The
work is also supported in part by PRIN--INAF 2014 ``Probing the AGN/galaxy
co-evolution through ultra-deep and ultra-high resolution radio surveys'', by
PRIN--INAF 2012 ``Looking into the dust-obscured phase of galaxy formation
through cosmic zoom lenses in the \textit{Herschel} Astrophysical Large Area
Survey'' and by ASI/INAF agreement n. 2014-024-R.1. M.B. and G.D.Z. acknowledge support from INAF under PRIN SKA/CTA FORECaST. Z.Y.C. is supported by the
National Science Foundation of China (grant No. 11503024). M.N. acknowledges
financial support from the European Union's Horizon 2020 research and
innovation programme under the Marie Sk{\l}odowska-Curie grant agreement No
707601.
\end{acknowledgements}

\begin{figure*}
\vskip-3cm
\centering
		\includegraphics[trim=0cm 0cm 0cm 0.cm,clip=true,width=0.99\textwidth,
    angle=0]{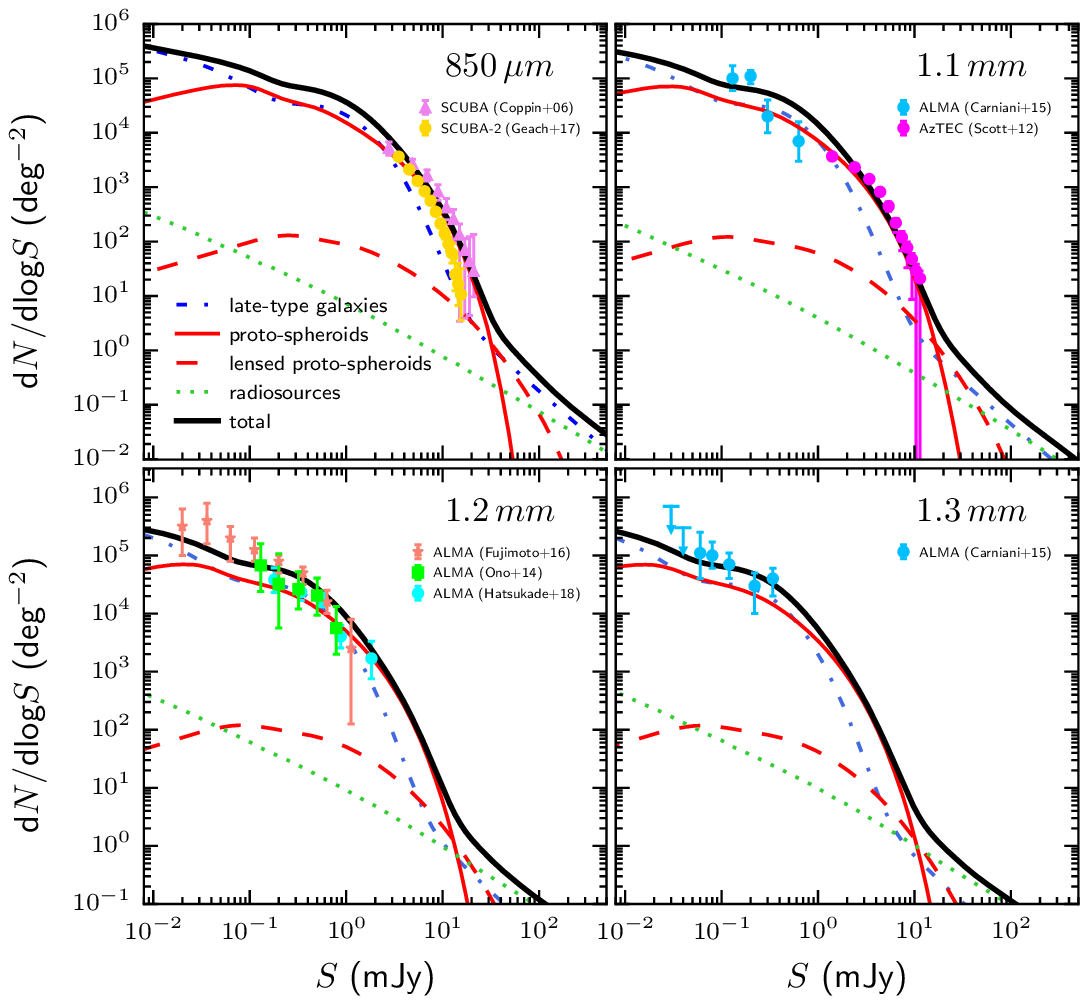}
  \caption{Comparison of the predictions of the \citet{Cai13} model with differential counts from
  recent SCUBA\,2 and ALMA surveys. The data are from \citet{Coppin2006}, \citet{Scott2012}, \citet{Ono2014},
  \citet{Carniani2015a}, \citet{Fujimoto2016} and \citet{Hatsukade2018}. }
  \label{fig:counts_new}
\end{figure*}

\begin{appendix}

\section{Model versus data}\label{sec:model_data}

As mentioned in Sect.~\ref{sect:intro}, our reference model was
validated against a broad variety of multifrequency data available until 2014.
Many additional data have accumulated since then. We have therefore decided to
check the model's consistency with these more recent data. Figure~\ref{fig:counts_new}
shows that the differential counts yielded by the model are in excellent
agreement with data from SCUBA\,2 and ALMA surveys. As illustrated by the
figure, the model includes contributions from different populations (normal and
star-bursting late-type galaxies, proto-spheroidal galaxies) and takes into
account the effect of gravitational lensing. The contribution of radio sources
to the counts is from \citet{Tucci2011}.

\begin{figure}
\centering
    \includegraphics[trim=3.2cm 1.0cm 1.5cm
    1.3cm,clip=true,width=0.49\textwidth, angle=0]{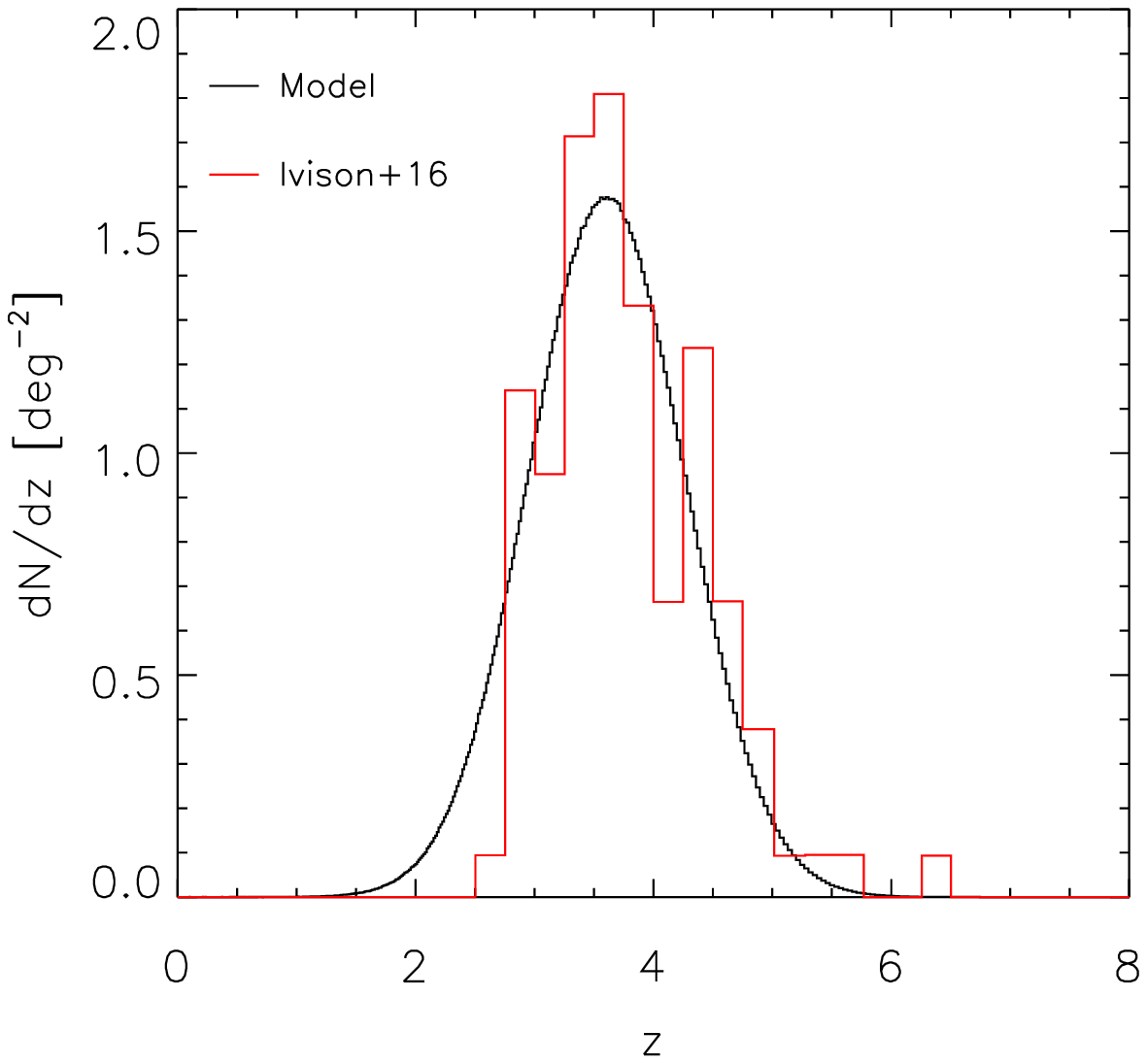}
\caption{Comparison between the \citet{Ivison2016} redshift distribution and the theoretical
one derived from the \citet{Cai13} model.}
 \label{fig:Ivison2016_comp}
\end{figure}

As mentioned in sub-section~\ref{sect:caveats}, recent estimates have
yielded space densities of $z\simgt 4$ dusty star forming galaxies well in
excess of model predictions. To check how our model behaves in this respect, we
have compared the predicted redshift distribution with the estimate by
\citet{Ivison2016}, based on a much larger sample than those previously available.
To this end, we simulated the selection criteria applied by
\citet{Ivison2016}: $S_{500\mu\rm m} \ge 30\,$mJy and ultra-red colours in the
\textit{Herschel}/SPIRE bands, (i.e., $S_{500\mu\rm m}$ larger than both
$S_{350\mu\rm m}$ and $S_{250\mu\rm m}$). Moreover, we took into account the
uncertainty in their photometric redshifts, $\Delta z/(1+z)=0.14$. To the estimate
of the total number of galaxies satisfying their selection criteria,
\citet{Ivison2016} applied a large correction for incompleteness (a factor of
$36.0\pm 8.2$); the result is therefore quite uncertain. Thus, we have chosen
to compare model predictions with the redshift distribution normalized to the
number of objects. As illustrated by Fig.~\ref{fig:Ivison2016_comp}, we find
good agreement with the shape of the distribution, although the normalization
is low by a factor of $\simeq 2.5$.

Furthermore, the model, coupled with the well established correlation between
radio and far-IR emission, successfully reproduces the radio luminosity
functions of SF galaxies, which is observationally determined up to $z\simeq 5$
\citep{Bonato2017radio}.

\end{appendix}

\bibliographystyle{pasa-mnras}
\bibliography{pasa}

\end{document}